\documentclass[a4paper,fleqn,usenatbib,useAMS]{mnras}
\usepackage{graphics}%%%,dblfloatfix}   % Including figure files
\usepackage{graphicx}%%%,dblfloatfix}   % Including figure files
\usepackage{amsmath}    % Advanced maths commands
\usepackage{amssymb}    % Extra maths symbols
\usepackage{multicol}        % Multi-column entries in tables
\usepackage{bm}     % Bold maths symbols, including upright Greek
\usepackage{pdflscape}  % Landscape pages
\usepackage{xcolor}
\usepackage{ae,aecompl}
\usepackage{subfigure}
\usepackage[english]{babel}
\usepackage{times}
\usepackage{physics}
\usepackage{etoolbox}
\makeatletter
\patchcmd\@combinedblfloats{\box\@outputbox}{\unvbox\@outputbox}{}{%
	\errmessage{\noexpand\@combinedblfloats could not be patched}%
}%
\makeatother

\title[Microlensing of radially pulsating stars]{Microlensing of radially pulsating stars}
\author[Sajadian et al.]{Sedighe Sajadian$~^{1}$\thanks{E-mail: s.sajadian@iut.ac.ir}, Richard Ignace$~^{2}$\thanks{E-mail: ignace@etsu.edu}
	                            \footnotemark[1]\\
	$^{1}$Department~of~Physics,~Isfahan~University~of~Technology,~Isfahan~84156-83111,~Iran\\
	$^{2}$ Department of Physics \& Astronomy, East Tennessee State 
	University, Johnson City, TN 37614, USA}

\date{Accepted XXX. Received YYY; in original form ZZZ}
\pubyear{2020}

\begin{document}
\label{firstpage}
\pagerange{\pageref{firstpage}--\pageref{lastpage}}
\maketitle
\begin{abstract}
	
Here, we study the microlensing of radially pulsating
stars. Discerning and characterizing the properties of
distant, faint pulsating stars is achievable through high-cadence
microlensing observations. Combining stellar variability
period with microlensing gives the source distance, type, and radius
and helps better determine the lens parameters. Considering
periodically variations in their radius and surface temperature,
their microlensing light curves are resulted from multiplication of the
magnification factor with variable finite size effect by the intrinsic
brightness curves of pulsing source. The variable finite source size
due to pulsation can be significant for transit and single microlensing
and while caustic-crossing features. This kind of deviation in the
magnification factor is considerable when the ratio of
the source radius to the projected lens-source distance is in the range
of $\rho_{\star}/u \in[0.4,10]$ and its duration is short and in the
same order of the time of crossing the source radius. Other deviations
due to variable source intensity and its area make colored and periodic
deviations which are asymmetric with respect to the signs of pulsation
phase. The positive phases makes deviations with larger amplitude that
negative phase. These deviations dominate in filters
with short wave lengths (e.g., $B-$band). The position of magnification
peaks in microlensing of variable stars varies and this displacement
differs in different filters.

\end{abstract}


%%%%%%%%%%%%%%%%%%%%%%%%%%%%%%%%%%%%%%%%%%
\section{Introduction}

Most stars display some variation of brightness with time; for instance,
even the Sun is a variable star, but its luminosity changes by only
around $0.1\%$ over the eleven year solar cycle \citep{Suncycle}. 
Variable stars can be classified into two basic categories: extrinsic variables
such as binary eclipsing stars \citep[see, e.g.,][]{eclipsing}, and
intrinsic variables such as pulsating variable stars \citep[see, e.g.,][]{zhevakin1963,cox1980,Pel1985,saio1993,Carrollbook}. The variation
in the brightness of radially pulsating stars arises from periodic
expansion and contraction, with consequent variations in the rate at
which radiation emerges from the stellar surface. These stars are mostly
bright giants or super-giants, much more luminous than the Sun,
and their periods range from days to months. They are classified
according to their periods of variation and the shape of their
light curves, as reflective of evolutionary stages and initial
mass \citep{book2015,book2007}. For radially pulsating stars, there is a relation between their luminosity and the
period of their variation, the so-called Leavitt's law, which makes
radial pulsators such as Cepheid and RR Lyrae stars standard candles
\citep[see, e.g.,][]{periodlumi,periodlumi2}. The relation between the
pulsation characteristics, stellar evolution and stellar properties,
along with their use as standard candles, makes the detection and detailed
study of radial pulsators important to astrophysics, even beyond stellar
astrophysics \citep[see, e.g.,][]{Cepheids}. For instance, the discovery
of Leavitt's Law and identification of Cepheids in other galaxies led
to the Hubble expansion law, which completely changed our perspective of the Universe 
\citep{Hubblelaw,Hubblelaw2}. To date, more than $55,000$ 
confirmed variable stars can be found in \textit{General Catalogue of Variable Stars (GCVS)}\footnote{\url{http://www.sai.msu.su/gcvs/gcvs/}} \citep[][]{samus2009,samus2017}.

Identifying pulsating stars requires intensive observations over several
decades \citep{OGLE_V1, OGLE_V2,variableN}. However,
some of variable stars, e.g., low-amplitude variables,
short period ones and irregular variables, are difficult
to identify and classify. Spectroscopic observations from
these puzzling variables can reveal their types and properties
\citep{lowamplitude,spec2015,Pawel2017}. But spectroscopy is helpful only
for bright source stars. Distant (and therefore faint), low-amplitude
variable stars can become observable through brightness magnification
during gravitational microlensing events. Macrolensing refers
to the ability to detect multiple images from gravitional lensing,
whereas microlensing refers to the limit when the multiple images are not
resolved, yet do lead to brightness variations. Macrolensing occurs
on cosmological scales, whereas microlensing refers to lensing masses
like stars for distance scales relevant to the Milky Way and Local
Group \citep{schneider1992,Liping2005}.  Currently, monitoring efforts
to detect gravitational microlensing events are conducted toward the
Galactic Bulge \citep{Mroz2019,bond2001,KMTNet}.  A
background source star is magnified when nearly aligned with a foreground
mass \citep{Einstein1936,pac86,wambsganss2006}. Microlensing has been
employed in recent years as a technique for detecting extra-solar
planets owing to how their masses modify stellar light curves from normal
lensing events associated with stellar masses
\citep{Mao1991,gould1992,Gaudi2012}.

The brightness magnification from microlensing can help with discerning
short and/or low-amplitude variations in the variable star light
curve which is intrinsically faint. Microlensing
is the only method for highlighting small stellar perturbations, even
time-variable ones \citep[see, e.g.,][]{earthmass}. Through magnifying
faint stars that are pulsators, we can better identify the pulsational period to
constrain the absolute magnitude of the source star and determine the
star's distance. Determining the source distance, 
its type, and radius independently in microlensing events helps better
constrain the mass and distance of the lens object.

Microlensing surveys, such as OGLE, MOA, and KMTNet, generally have not
emphasized pulsating variables as source stars of microlensing events
\citep{RemoveVar}. However, one example of microlensing of a pulsating
star was discovered recently \citep{Varmicro}. For this
microlensing event, $\rm{OGLE}$-$\rm{2017}$-$\rm{BLG}$-$\rm{1186}$, the
source star was a bright and variable red giant. Its amplitude
of variation was high enough so that asteroseismology from ground-based
data revealed the pulsation curve and as a result its stellar type,
average source size, and distance. Measuring the finite source
size \citep{Gould1994,1994wittmoa,Nemiroff1994} carefully in addition to the parallax effect provides a solution
for the mass
and distance of the lensing star. However, the analysis of this
event did not take account of the variation
in the physical extent of the star.  Clearly, radial pulsations 
mean the source radius is not fixed.
Here, we investigate the effect of varying finite source size for
the lensing magnification.

In this paper, Section~\ref{two} provides a formalism for microlensing
of radially pulsating stars. There we introduce an overall magnification
factor as depending on the time varying size and temperature of the star
along with how the traditional magnification factor is averaged of the
finite stellar size.  We explore the contributions of these different
effects.  In Section \ref{three} results for single-lens events are
presented and discussed.  Then results for the binary-lens scenario are
given in Section \ref{four}.  We summarize these results and provide
concluding remarks in the last section.

\begin{figure}
	\centering
	\includegraphics[width=0.49\textwidth]{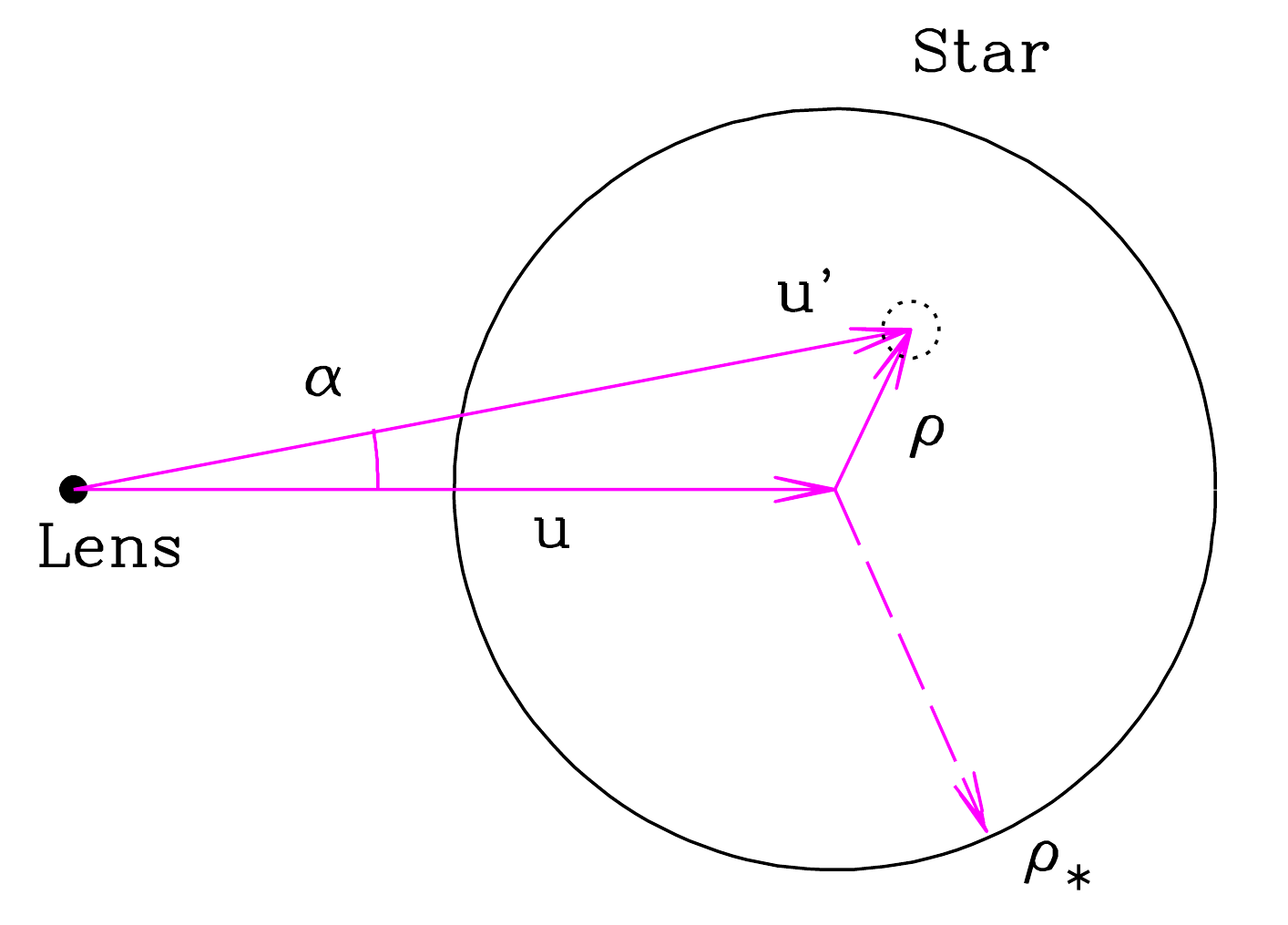}
	\caption{Figure represents the lens plane which contains the projected source surface with the radius $\rho_{\star}$. The coordinates of the source center and an arbitrary point of it with respect to th lens location is specified as $(u,0)$ and $(u' \cos \alpha, u' \sin \alpha)$, respectively.}\label{geomee}
\end{figure}
%%%%%%%%%%%%%%%%%%%%%%%%%%%%%%%%%%%%%%%%%%%%%%%%%%%%%
\section{Formalism for microlensing of radial pulsating stars} \label{two}
Radial pulsating stars vary in both effective surface temperature and
radius leading to periodic modulations of their observed specific fluxes
\citep{cox1980}. While the flux and effective temperature
curves have similar phases versus time, variations in the radial extend
of the atmosphere will lag these by around quarter in pulsational phase.
Since our objective is to explore basic observable consequences from
the microlensing of pulsating stars, we adopt a somewhat simplistic for
atmospheric variations of a radial pulsating star.  We choose periodically
sinusoidal functions for variations of the stellar radius, $R_\ast$,
and effective stellar temperature, $T_\ast$, as follows \citep{Helen}:

\begin{eqnarray}
R_{\star}(t)&=& \bar{R} + \delta_{R} \sin[\omega\,(t-t_{p}) +\phi_{0}], \, {\rm and}\\
T_{\star}(t)&=& \bar{T} + \delta_{T} \sin[ \omega\,(t-t_{p}) ],
\end{eqnarray}

\begin{figure*}
	\centering
	\includegraphics[width=0.49\textwidth]{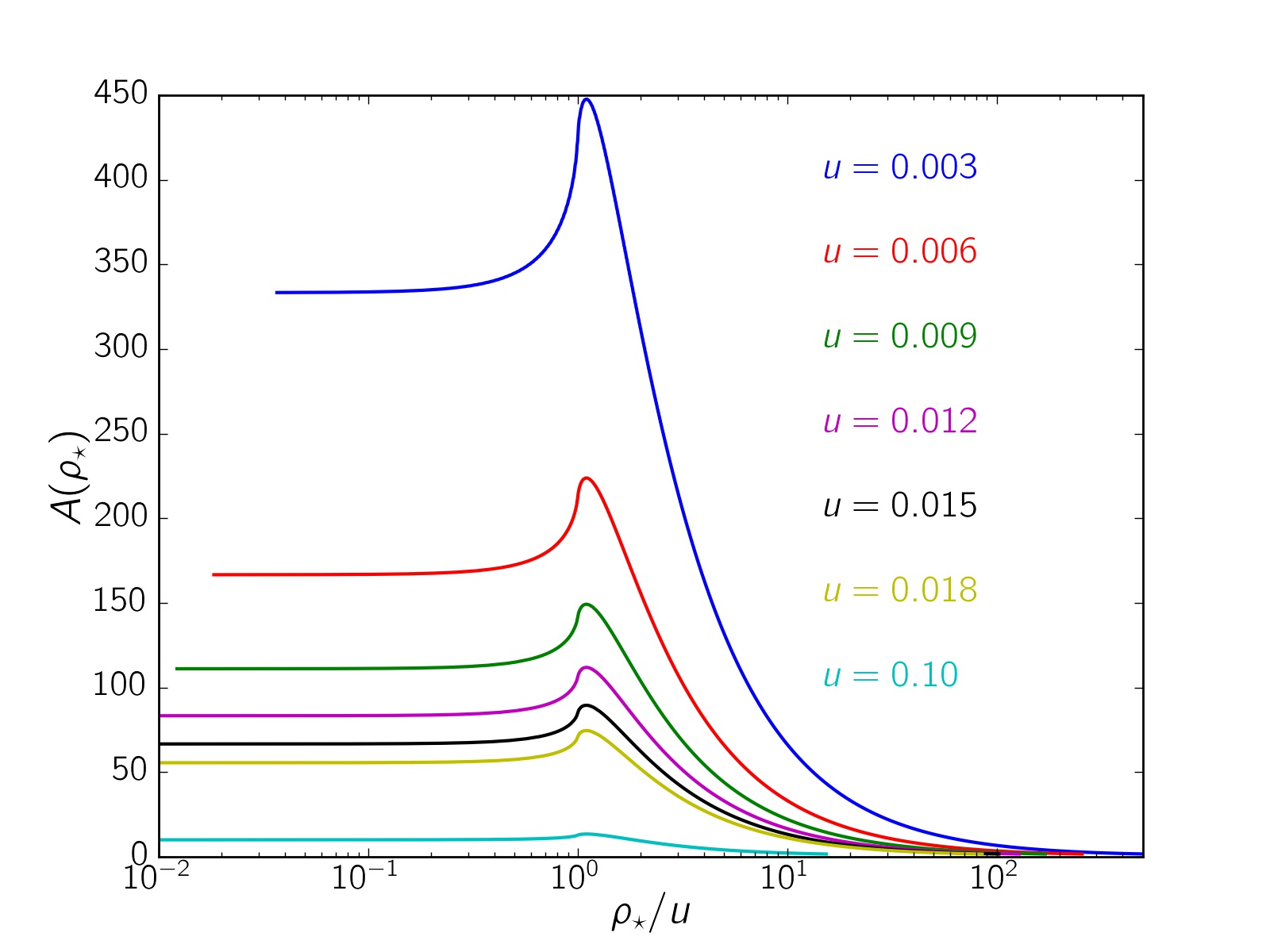}
	\includegraphics[width=0.49\textwidth]{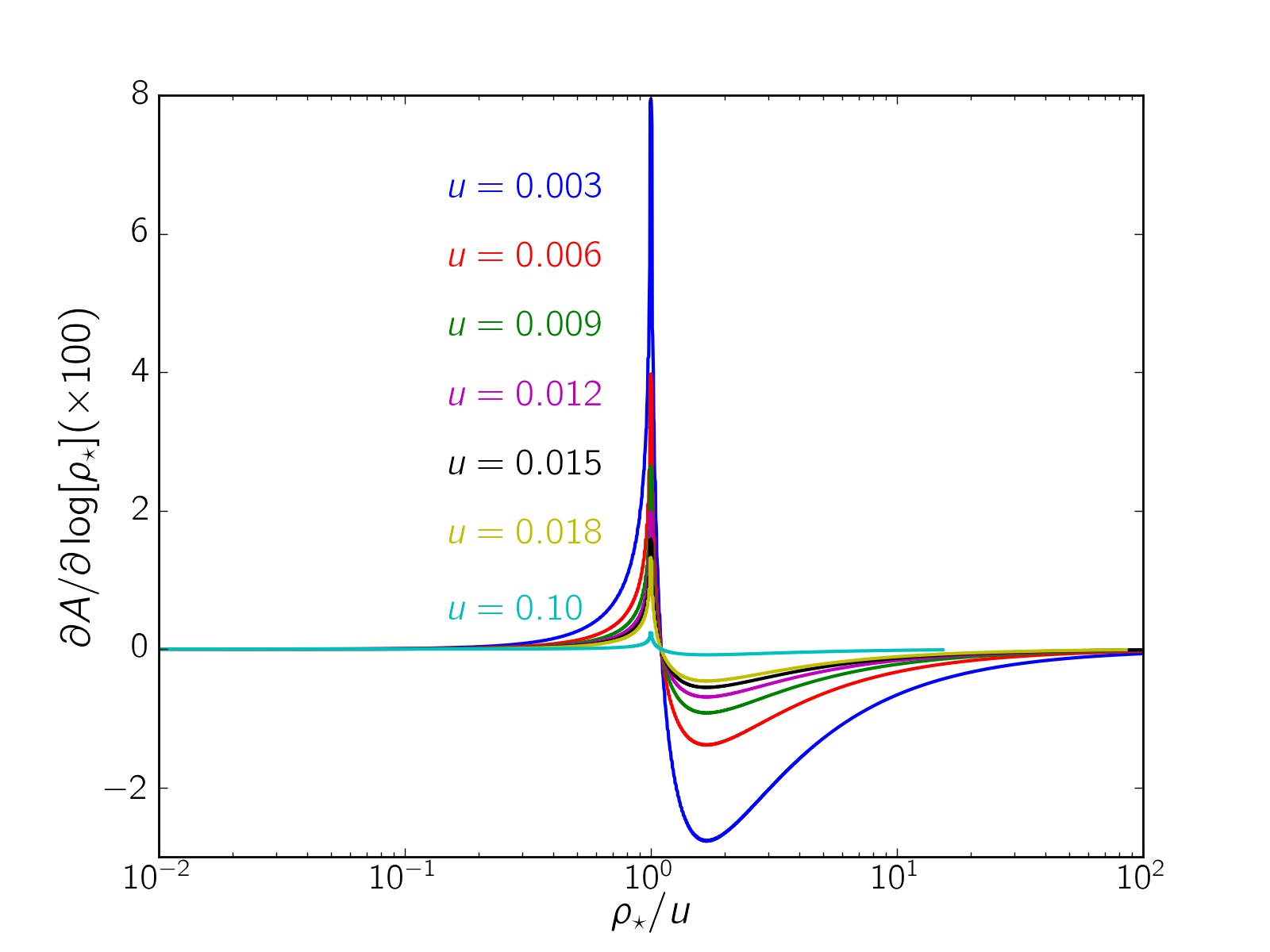}
	\caption{The microlensing magnification factor $A(\rho_{\star},u)$ (left panel) and its derivative (right panel) versus $\rho_{\star}/ u$ for
		different values of $u$, the lens distance from the source center normalized to the Einstein radius.}\label{delta1}
\end{figure*}
\noindent where $\bar{R}$ and $\bar{T}$ are the average radius and
effective temperature of the source star of one pulsational cycle,
$t_{\textrm{p}}$ is an arbitrary time offset, $\omega= 2 \pi /\textrm{P}$
is the pulsating angular velocity with $\textrm{P}$ the period of
pulsation, and $\phi_{0}$ is the phase difference between the radius and
temperature curves with $\phi_0\approx -\pi/2$. Then $\delta_{\textrm{R}}$
and $\delta_{\textrm{T}}$ are relative amplitude factors for the variation
of radius and temperature, respectively.

We further treat the star as a a blackbody for the specific intensity
of emission from the atmosphere, with $B_{\lambda}(T_{\star})$:

\begin{equation}
B_{\lambda}(T_{\star})=\frac{8\pi h c^{2}}{\lambda^{5}}\frac{1}{\exp(\frac{hc}{kT_{\star} \lambda})-1},
\end{equation}
where $\lambda$ is the wavelength, $c$ is the speed
of light,  $k$ is the Boltzmann constant,  $h$ is the Planck constant, and $B_{\lambda}$ is given in units of $\rm{W.m^{-3}}$.  \noindent The time-dependent passband flux of the source star 
received by the observer becomes,

\begin{eqnarray}\label{Fstar}
F_{\star}(t)=\left[\frac{R_{\star}(t)}{D_{s}}\right]^{2}\, \times
	\pi\,I_{\star}(t),
\end{eqnarray}

\noindent where $D_{\rm s}$ is the distance to the source star, and the passband dependence enters through

\begin{equation}
I_{\star}(t)=\int_{0}^{\infty} d\lambda \, K(\lambda-\lambda_0) \,
B_{\lambda}(T_{\star}).
	\label{eq:Istar}
\end{equation}

\noindent This ``passband intensity'' emitted from the stellar surface
is associated with an observational filter characterized by wavelength
$\lambda_0$.  The throughput of the filter, $K(\lambda-\lambda_0)$,
is taken to have unit area when integrated over wavelength.  Note that
in keeping with the theme of a simplified model, the star is taken to
have no limb darkening. However, equation~(\ref{eq:Istar}) could be
modified to include the effect of limb darkening.

To evaluate the amplification factor from microlensing of
a radial pulsating star, Figure (\ref{geomee}) illustrates the geometry in
the plane of the sky.  The star is taken to have angular radius
$\theta_\ast = R_\ast(t)/D_{\rm s}$.  However, it is convenient to express angular 
separations relative to the angular Einstein radius, the latter
being $\theta_{\rm{E}} = R_{\rm{E}} / D_l$, where $R_{\rm{E}}$ is the linear Einstein
radius in the lens plane, and $D_l$ is the distance to the lens.
A stellar radius parameter introduces as

\begin{equation}
\rho_\ast = \theta_\ast / \theta_{\rm{E}}.
\end{equation}

\noindent Related, we define $\theta$ as the angular separation
in the sky between the lens and the center of star, and we introduce

\begin{equation}
u = \theta / \theta_{\rm{E}}= \sqrt{u_{0}^{2} +\left(\frac{t-t_{0}}{t_{\rm{E}}}
	\right)^{2}},
\end{equation}
\noindent 
as the lens-source separation normalized to the
Einstein radius and projected on the lens plane.  Here, $u_{0}$,
is the minimum separation (i.e., impact parameter) between the lens
and source on the lens plane and normalized to the Einstein radius.
Correspondingly, $t_{0}$ is the time of closest approach when $u_0$
is achieved.  The Einstein crossing time, $t_{\rm{E}}= R_{\rm{E}}/v_{t}$,
is the time to cross the Einstein radius given the lens-source transverse
velocity $v_{t}$. If the source radius is negligible in comparison with
$u$, i.e., $u\gg \rho_\ast$, the magnification factor for the single
lens is given by:
\begin{eqnarray}\label{magnii}
A(u)=\frac{u^{2}+2}{u\sqrt{u^2+4}}.
\end{eqnarray}
The curve of this magnification factor versus time is known as the simple Pczy\'nski light curve \citep{pac86,1994wittmoa}.

\noindent For microlensing that takes account of the finite stellar size, we define $\theta '$ and $u' = \theta ' / \theta_E$ which are dummy variables for determining the amplification factor $A(u,\rho_\ast)$, with\\
\begin{eqnarray}
A(u,\rho_\ast) = \frac{\int A(u')\,I(u',\alpha)\,u'\,du'\,d\alpha}
	{\int I(u')\,u'\,du'\,d\alpha},
\end{eqnarray}

\noindent and $A$ is given by Equation (\ref{magnii}).
Note that in our treatment that currently ignores either limb
darkening or non-radial pulsations, $I(u',\alpha) = I_\ast (t)$,
and the above expression simplifies to

\begin{equation}
A(u,\rho_\ast) = \frac{1}{\pi\,\rho_\ast^2}\,
	\int A(u')\,u'\,du'\,d\alpha.
\end{equation}

Note that $A(u,\rho_\ast)\ge 1$ has no units owing to normalization by the
unlensed brightness of the star.  The case of $A=1$ represents no lensing.
The case of $u\gg \rho_\ast$ is the traditional case of point lensing
of a point-like star. But when $u \lesssim \rho_\ast$, the amplification
of the star changes relative to treating the star as a point source.
While $A$ is implicitly time-dependent by virtue of this surface averaging
over a star whose extent is varying, we need another factor to represent
the {\em observed} amplification of the source, which we define as

\begin{equation}\label{magn}
A_{o}(t) \equiv \frac{I_{\star}(t)}{\bar{I}}\left[\frac{R_{\star}(t)}
	{\bar{R}}\right]^{2}A(u,\rho_{\star}),
\end{equation}

\noindent where the time-averaged passband intensity is given by
\begin{equation}
\bar{I} = \frac{1}{P}\,\int_0^P\,I_\ast(t)\,dt.
\end{equation}

\noindent The intrinsic variation in the stellar temperature and its
radius makes time-dependent perturbations in the magnification factor.

\begin{figure}
\centering
\includegraphics[width=0.49\textwidth]{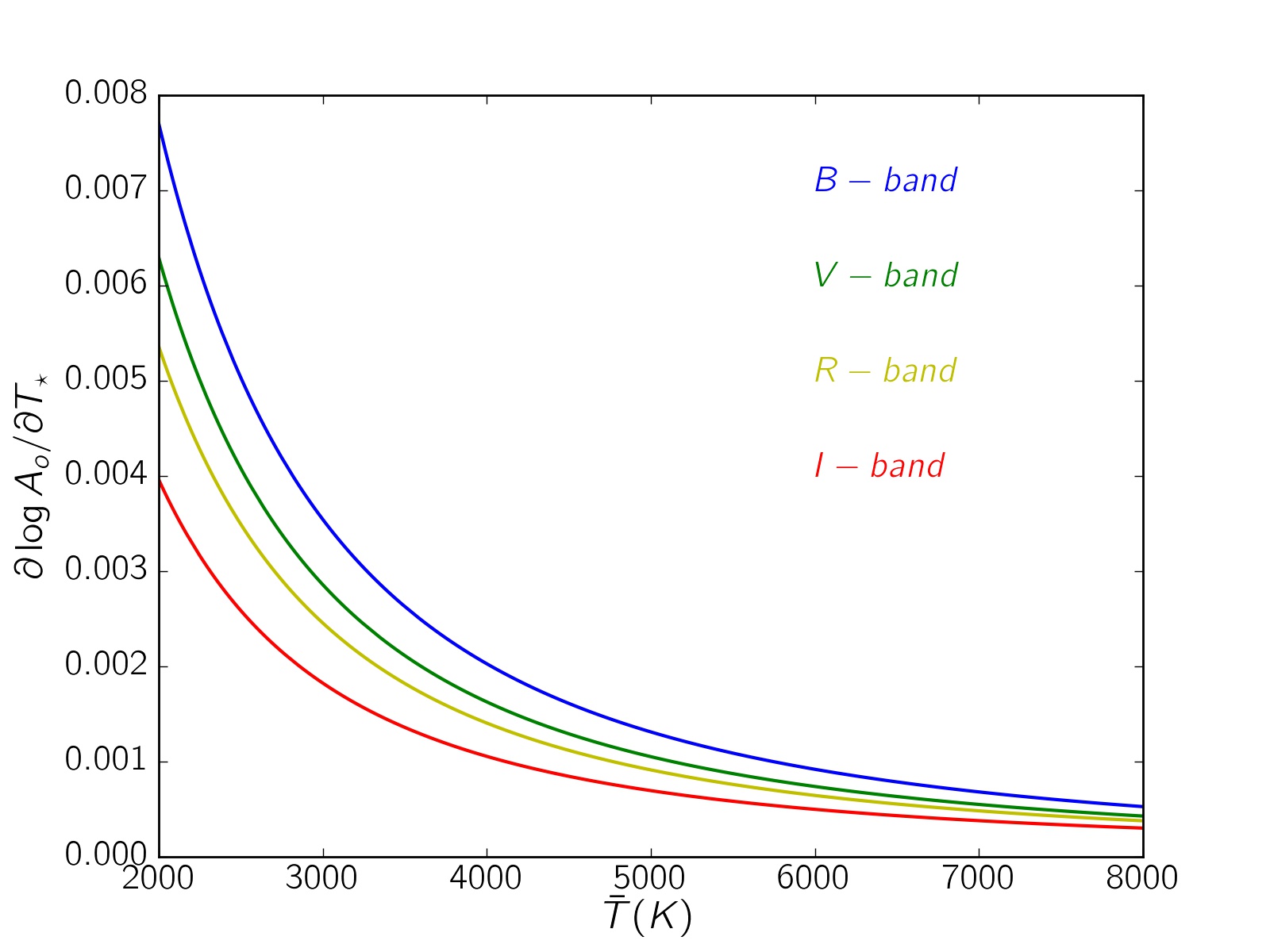}
\caption{The partial derivatives of the observed magnification factor $A_{o}$ with respect to $T_{\star}$ which is normalized to the  observed magnification factor versus the source surface temperature.}\label{delta3}
\end{figure}

%%%%%%%%%%%%%%%%%%%%%%%%%%%%%%%%%%%%%%%%%%%%%%%%%%%%%%%%%%%%%%
\subsection{General characterizations}
According to equation (\ref{magn}), there are three sources of
time-dependent variations associated with the microlensing of radial
pulsators (i.e., in addition to the relative motion between the lens
and source): (i) the finite source size $\rho_{\star}$ which changes the
magnification factor, (ii) the source radius which changes the emitting
area of the source by $(R_{\star}(t) / \bar{R})^{2}$, and (iii) the
intrinsic intensity of the source star owing to the variation in the
stellar temperature $I_{\star}(t)/\bar{I_{\star}}$.  As a result, the
observed microlensing light curve will deviate from a standard Paczy\'nski light curve (c.f., Eqn~\ref{magnii}). 
To understand when the pulsation-induced deviation
in the observed magnification factor becomes significant, we investigate
each of these 3 factors separately.  For this initial study, we
focus on the single-lens case.

$\mathbf{(i)~A(\rho_{\star}(t),u)}$: The lefthand side of
Figure~(\ref{delta1}) shows the
magnification factor $A$ as plotted against $\rho_{\star}/u$ for several
values of lens-star separations, $u$. Maxima are found numerically to
occur when $u=\rho_{\star}/1.1=0.91\rho_{\star}$.  This corresponds
to where the projected position of the lens is slightly interior to the
projected source surface.  We calculated the magnification factor with
finite source size using the $\rm{RT}$-model developed by V.\ Bozza
\citep{Bozza2018,Bozza2010,Skowron2012}.  The magnification factor
decreases for $\rho_{\star}/u>1.1$ and the calculation approaches that of
the point-lens with point-source as the finite size of the star becomes
diminishingly relevant.  For $\rho_{\star}/u <1.1$, the magnification
factor decreases somewhat but then approaches a near-constant value as
$\rho_{\star}\rightarrow 0$.

To investigate the rate of variation of the magnification factor due to
the variation in $\rho_{\star}$, we differentiate the magnification factor
with respect to $\rho_{\star}$ and plot the result in the rightside
panel of Figure~(\ref{delta1}).  The sharp positive peaks of $\partial A/
\partial \ln \rho_{\star}$ occur when $\rho_{\star}=u$ (turning point),
and the derivative changes sign when $\rho_{\star}=1.1 u$. If the lens
distance from the source center is larger than $\rho_{\star}/1.1$,
this derivative, $\partial A/\partial \rho_{\star}$, is positive, such
that increasing the source radius enhance the magnification factor.
Contrastly, when $u<\rho_{\star}/1.1$, the derivative is negative. In
conclusion, the magnification factor displays its greatest response to the
variable size of the star when $\rho_{\star}\sim u$.\\

The derivative normalized to the magnification factor is related to
the relative deviation in the observed magnification factor as:

\begin{eqnarray}\label{del1}
\frac{\partial \ln A_{o}}{\partial \rho_{\star}}= 
	\frac{\partial \ln A}{\partial \rho_{\star}}.
\end{eqnarray}

\noindent This relative deviation in $A_{o}$ is small when
the averaged source radius is larger than $\sim 10 u$ and less than
$0.4 u$. Hence, the considerable deviation happens when $0.4\lesssim
\rho_{\star}/u \lesssim 10$. The time scale of this deviation is in the
same order of magnitude of $t_{\star}=t_{\rm{E}}\rho_{\star}$, the time of
crossing the source radius. In order to estimate the value of $t_{\star}$,
consider as an example a microlensing event toward the Galactic
Bulge with typical parameters: a Bulge distance $D_{s}=8.5~$kpc, a lens
distance $D_{l}=6~$kpc, a lens mass $M_{l}=0.3M_{\odot}$,
a transverse velocity $v_{\rm{t}}=120~$km/s, 
a Bulge giant star radius $R_{\star}=10R_{\odot}$ which results in
$R_{\rm{E}}=2.1~$au,  $t_{\rm{E}}=29.9~$days, and $\rho_{\star}=0.02$.
These parameters then yield $t_{\star}=0.5~$day. During this time scale, measuring the pulsation
period is impossible, unless $p\lesssim t_{\star}$ (which is rare).

\begin{figure}
	\centering
	\includegraphics[width=0.49\textwidth]{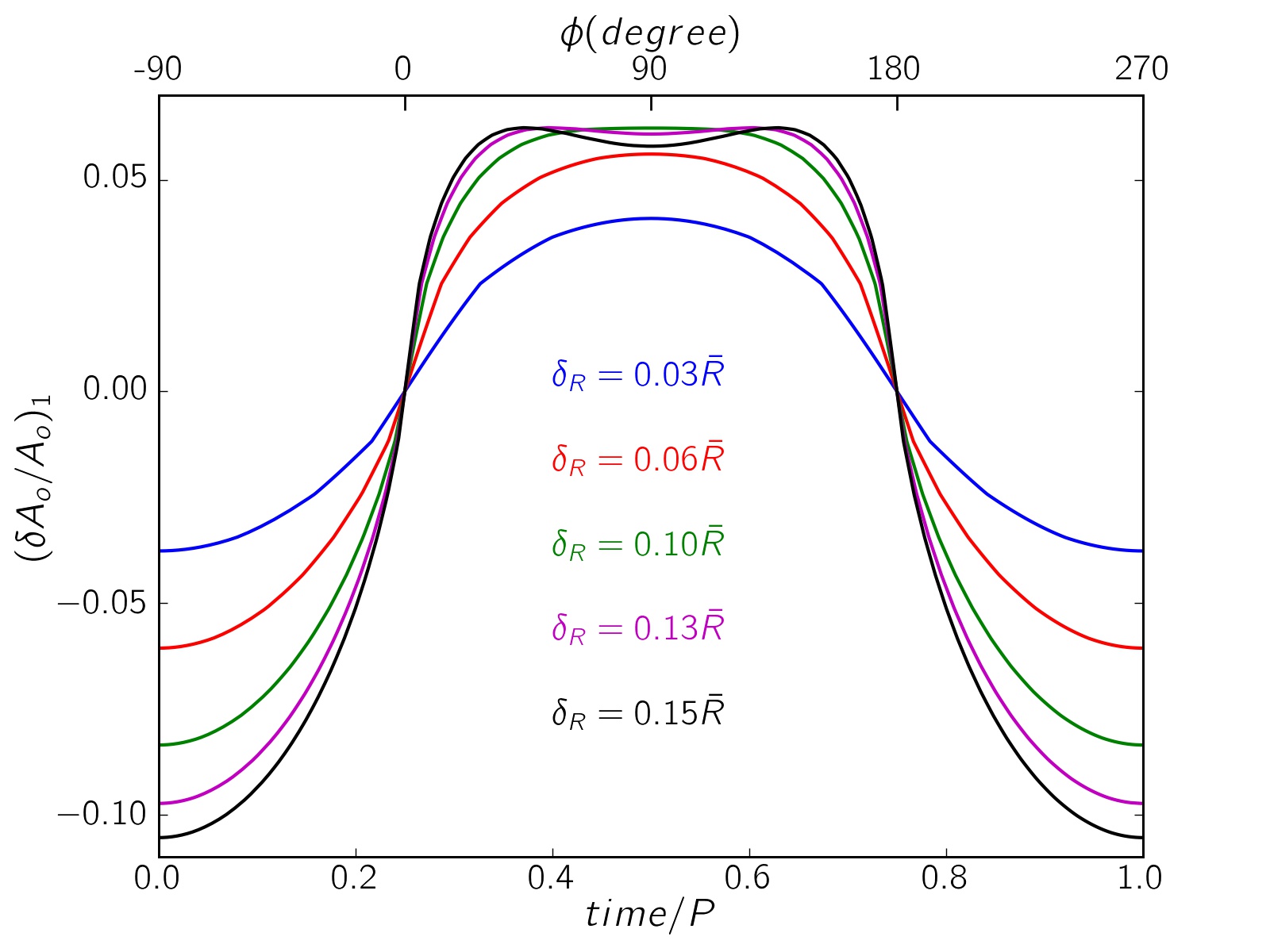}
	\includegraphics[width=0.49\textwidth]{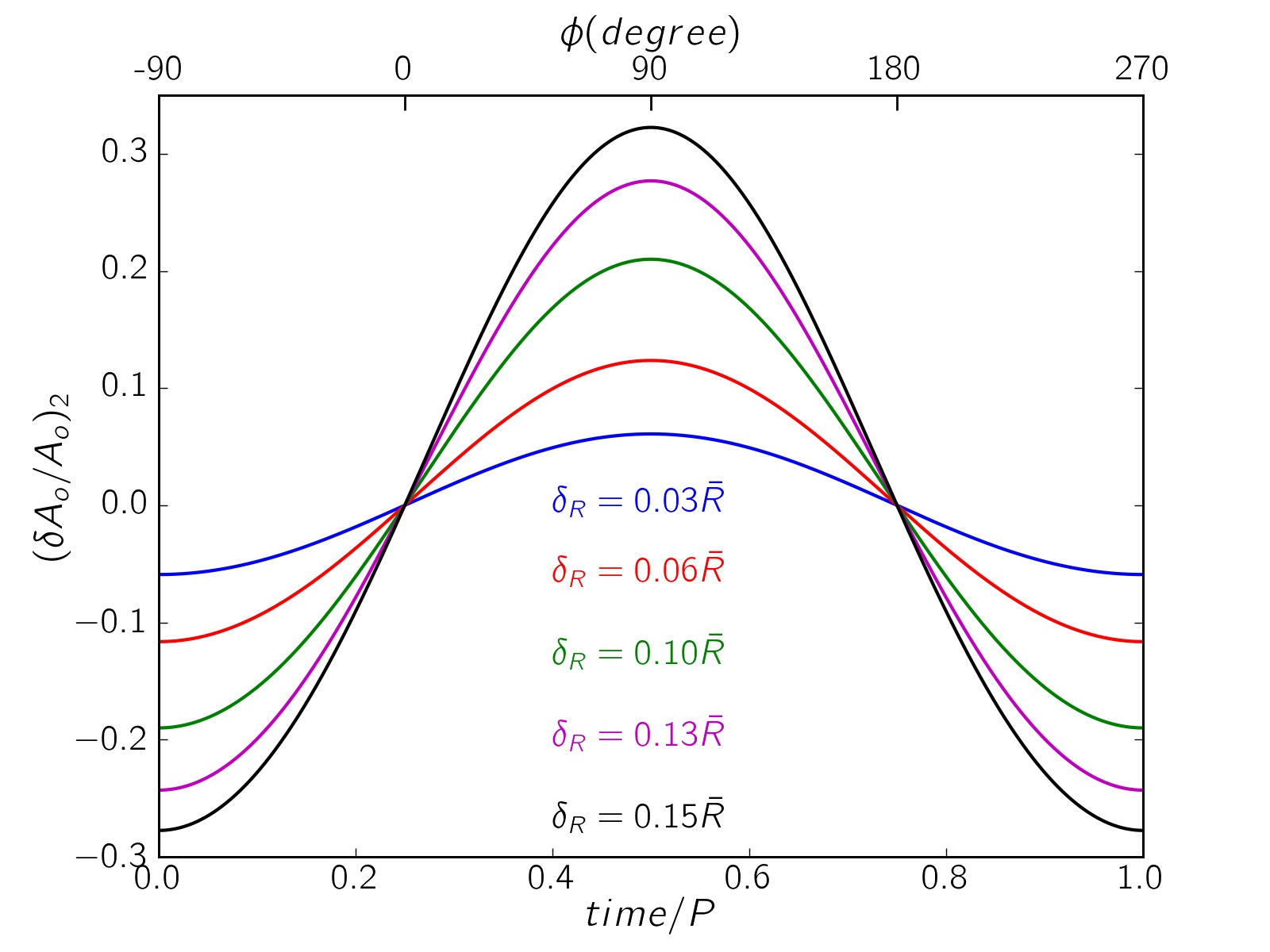}
	\includegraphics[width=0.49\textwidth]{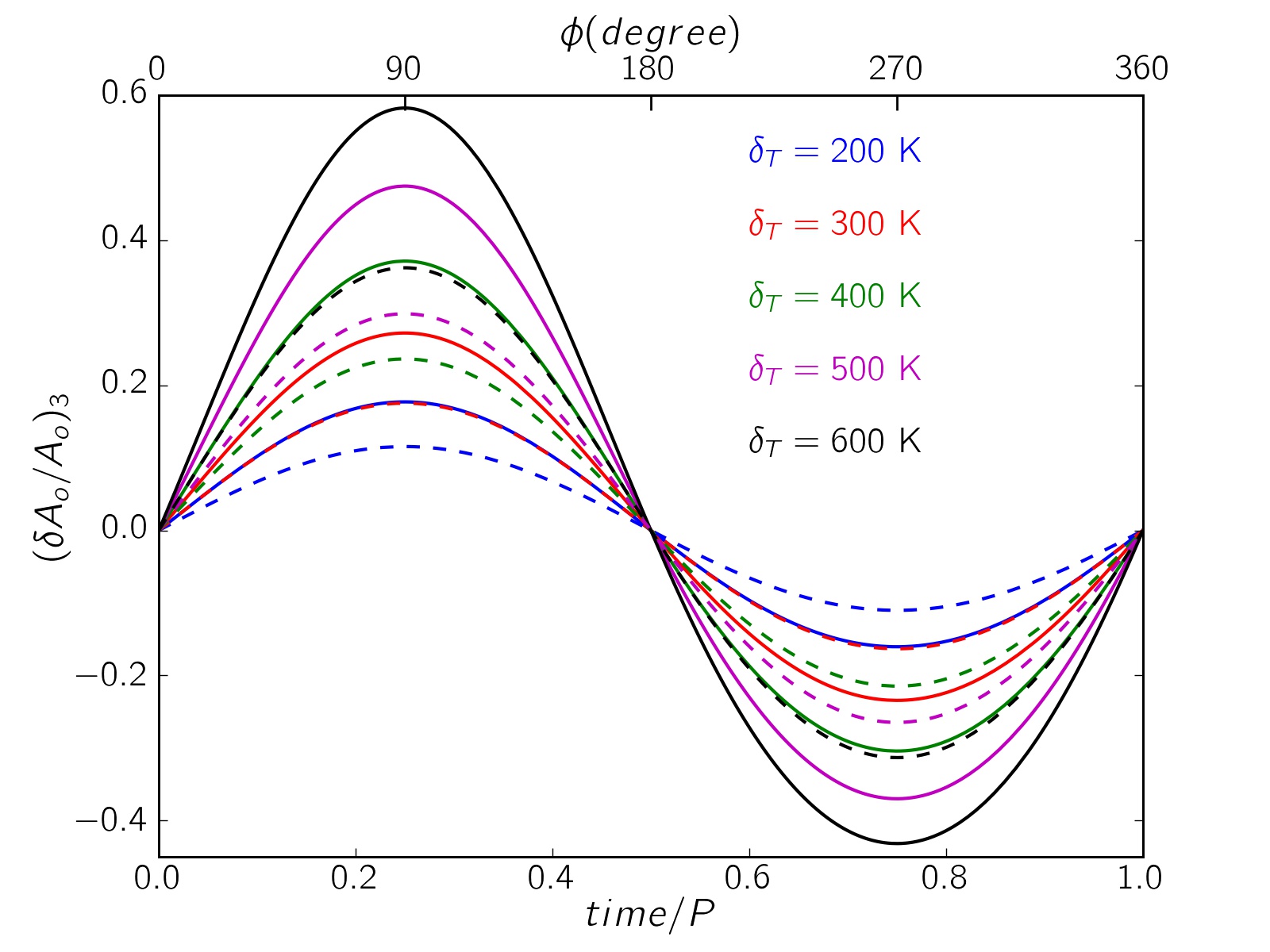}
	\caption{The relative deviations in the observed magnification factor due to the deviation in $A$ (top plot),  $R_{\star}^{2}$ (middle plot) and $I_{\star}$ (bottom plot) during one period  for different amplitudes $\delta_{R}$ and $\delta_{T}$. In the last plot, the solid and dashed curves are in filters $V$ and $I-$bands.}\label{phi}
\end{figure}

$\mathbf{(ii)~R_{\star}^{2}}$: The area of pulsating source stars changes
with time which makes variable intrinsic luminosity for the source
star. The relative derivative of the observed magnification factor with
respect to the source radius is given by:

\begin{eqnarray}\label{del2}
\frac{\partial \ln A_{o}}{\partial R_{\star}}= + \frac{2}{R_{\star}}. 
\end{eqnarray}

\noindent When the source radius is large, the relative rate of
variation in the magnification is small, and vice versa. Also, $\partial
A_{o}/ \partial R_{\star}$ is positive always, i.e., by increasing
$R_{\star}$ the observed magnification factor 
increases and vice versa. For stars of relatively small angular size,
the derivative of the observed magnification factor with respect to
$R_{\star}$ is large, but its derivative with respect to $\rho_{\star}$
is large only when $u\sim \rho_{\star}$. \\

$\mathbf{(iii)~I_{\star}(t)}$ The intrinsic intensity from pulsating source stars derives its variability from the
surface temperature. The relative derivative of the magnification with respect to the stellar surface
temperature is:

\begin{eqnarray}\label{del3}
\frac{\partial \ln A_{o}}{\partial T_{\star}}=\frac{\partial \ln I_{\star}}
	{\partial T_{\star}}.
\end{eqnarray}

\noindent This function is plotted in Figure~(\ref{delta3}) against
stellar temperature in different standard Johnson filters
$\textrm{BVRI}$. The variation in the stellar surface temperature
makes the observed magnification factor change differently in various
filters. The largest relative variations in the observed magnification
factor due to variation in the stellar temperature happens in
the $B$-band. For hotter stars the $\textrm{BVRI}$ are more in
the Rayleigh-Jeans tail, for which the above equation approaches
a $T_\ast^{-1}$ dependence. On the Wien side for quite cool stars,
the normalized derivative approaches $T_\ast^{-2}$, which is much
steeper. Hence, temperature variability leads to stronger effects in the magnification factor for cooler stars.

Generally, $T_{\star},~R_{\star}, \rm{and~} \rho_{\star}$ all change with
time, simultaneously, but not necessarily in phase. The relative rate
of variation in the observed magnification factor with time can be written as
a summation of its derivatives involving equations (\ref{del1}), (\ref{del2})
and (\ref{del3}) as follows:

\begin{eqnarray}
\frac{\partial \ln A_{o}}{\partial t}&=& \left[ \frac{\partial \ln A}{\partial \rho_{\star}}\,\delta_{\rho}+\frac{2}{R_{\star}}\delta_{R}\right] \, \omega \cos [\omega(t-t_{p})+\phi_{0}] \nonumber\\&+& \frac{\partial \ln I_{\star}}{\partial T_{\star}}\,\delta_{T}\,\omega\, \cos [\omega (t-t_{p})],
\end{eqnarray}

\noindent where $\delta_{\rho}=(\delta_{R} D_{l})/ (R_{\textrm{E}}D_{s})$
is $\delta_{R}$ projected onto the lens plane and normalized to the
Einstein radius. Accordingly, the derivative (the rate of perturbations)
in the observed magnification factor increases linearly with $\omega$,
$\delta_{R}$ and $\delta_{T}$ in addition to the mentioned effects.
Note that the sign of the relative derivative in the observed magnification
factor depends on the variation phase.

%%%%%%%%%%%%%%%%%%%%%%%%%%%%%%%%%%%%%%%%%%%%%%%%%%%%%%%%%%%%%%%%%%
\subsection{Small and periodic perturbations}

For the small variations in the source radius and its temperature,
we can use the Taylor expansions in time
around their average values as following:

\begin{eqnarray}
A(u,\rho_{\star})&=&A(u,\bar{\rho}_{\star}) + A^{\prime}(t) \times [\rho_{\star}(t)-\bar{\rho}_{\star}],\nonumber\\
R_{\star}&=&\bar{R} + R_{\star}^{\prime}(t) \times [R_{\star}(t)-\bar{R}],\nonumber \\
I_{\star}&=&\bar{I} + I_{\star}^{\prime}(t) \times [I_{\star}(t)-\bar{I}].
\end{eqnarray}

\noindent By substituting these expansions into equation~(\ref{magn}),
and considering only first-order terms, the relative deviation in the
observed magnification factor becomes:

\begin{eqnarray}
\frac{\delta A_{o}}{A_{o}}=  \frac{\delta A}{A} + \frac{\delta R_{\star}^{2}}{R_{\star}^{2}} + \frac{\delta I_{\star}}{I_{\star}}=\left(\frac{\delta A_{o}}{A_{o}}\right)_{1}+ \left(\frac{\delta A_{o}}{A_{o}}\right)_{2}+\left(\frac{\delta A_{o}}{A_{o}}\right)_{3}. 
\end{eqnarray}

\noindent The relative variation in the observed magnification
factor is the sum of these factors for small amplitudes of
variations. These relative variations are plotted in the three panels
of Figure~(\ref{phi}) versus time for one pulsation period. The fixed
parameters selected for these plots are $u=0.007$, $\bar{R}=5R_{\odot}$,
$\bar{\rho}_{\star}=0.0069$, and $\bar{T}=5600$~K. In the bottom plot,
the solid and dashed curves are for the $V$ and $I$-band filters,
respectively.
These variations are calculated with respect to average values (at
the baseline) and so they are not derivatives. Hence, the scales on
$y-$axes of the different panels can be compared directly for
how the different factors contribute to variability.

\begin{figure*}
	\centering
	\subfigure[]{\includegraphics[angle=0,width=0.49\textwidth,clip=0]{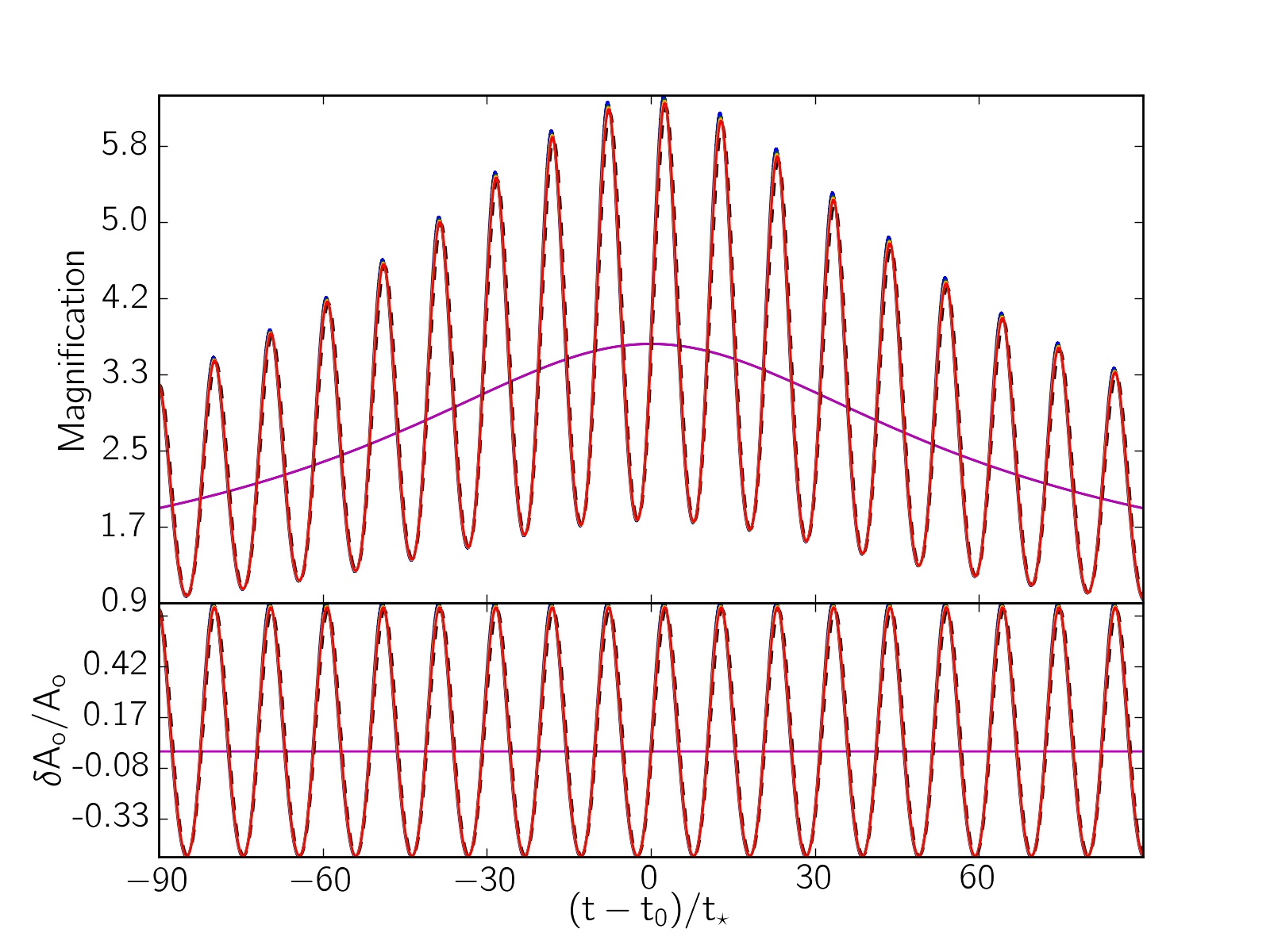}\label{figa}}
	\subfigure[]{\includegraphics[angle=0,width=0.49\textwidth,clip=0]{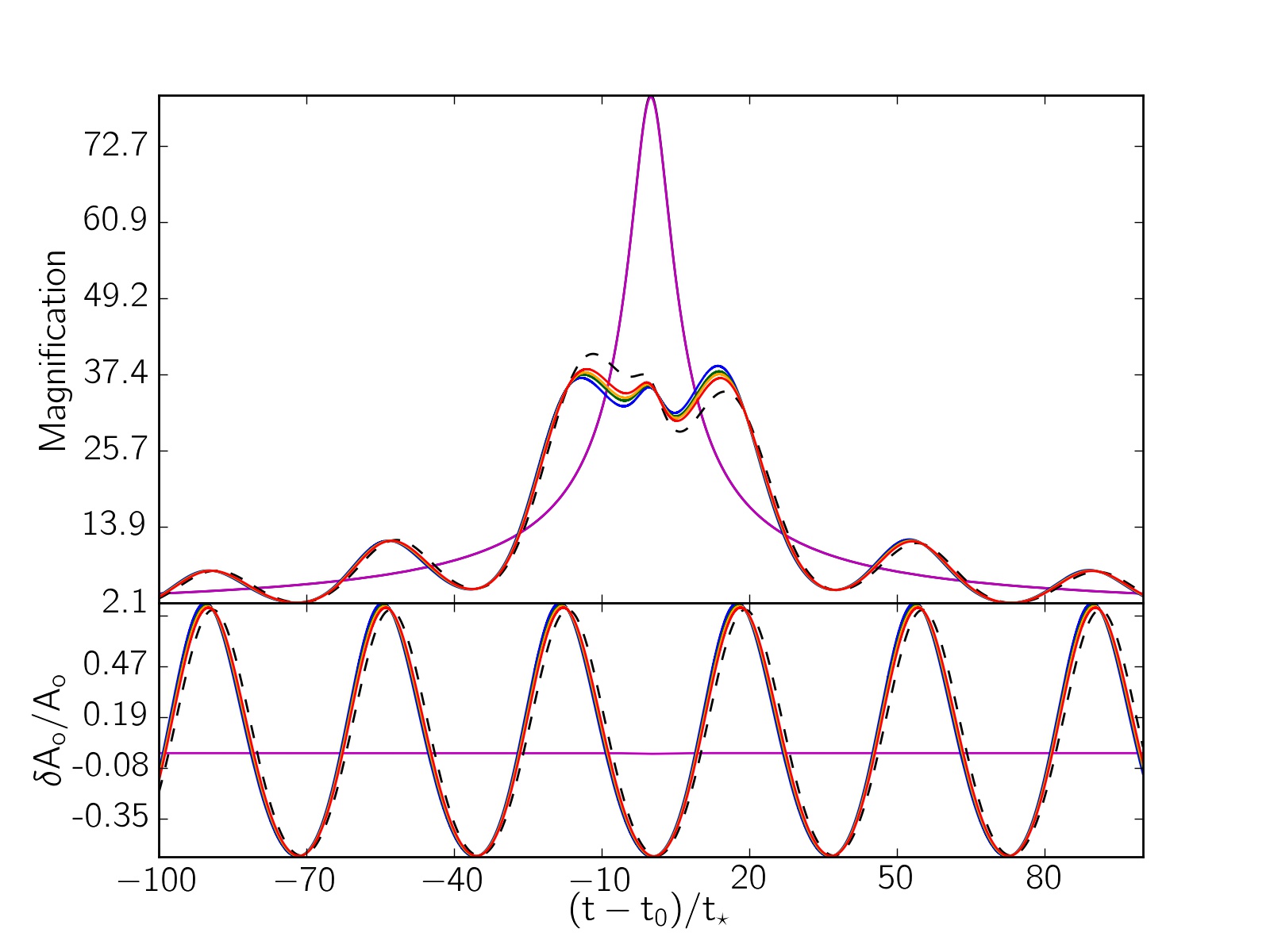}\label{figb}}
	\subfigure[]{\includegraphics[angle=0,width=0.49\textwidth,clip=0]{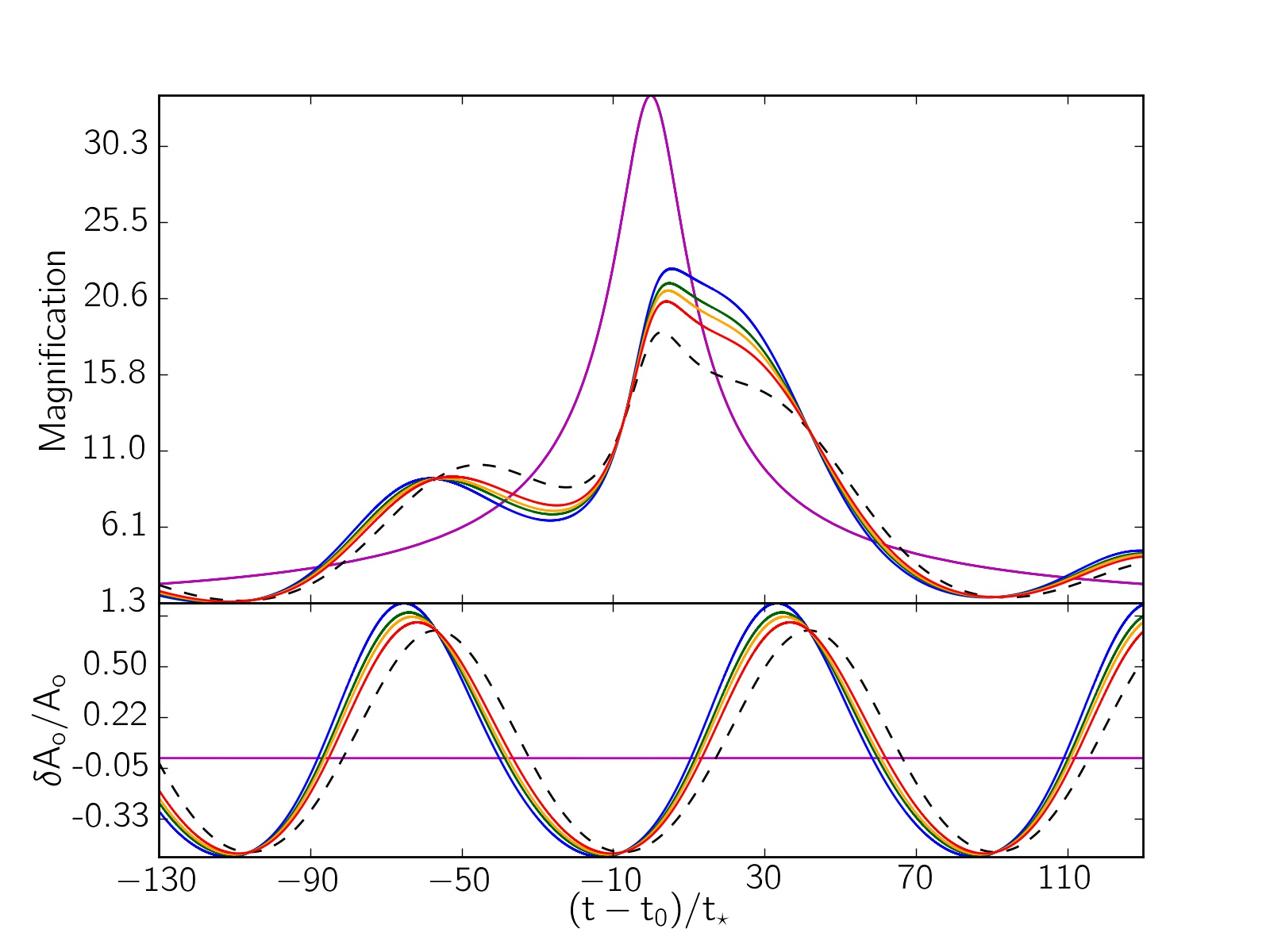}\label{figc}}
	\subfigure[]{\includegraphics[angle=0,width=0.49\textwidth,clip=0]{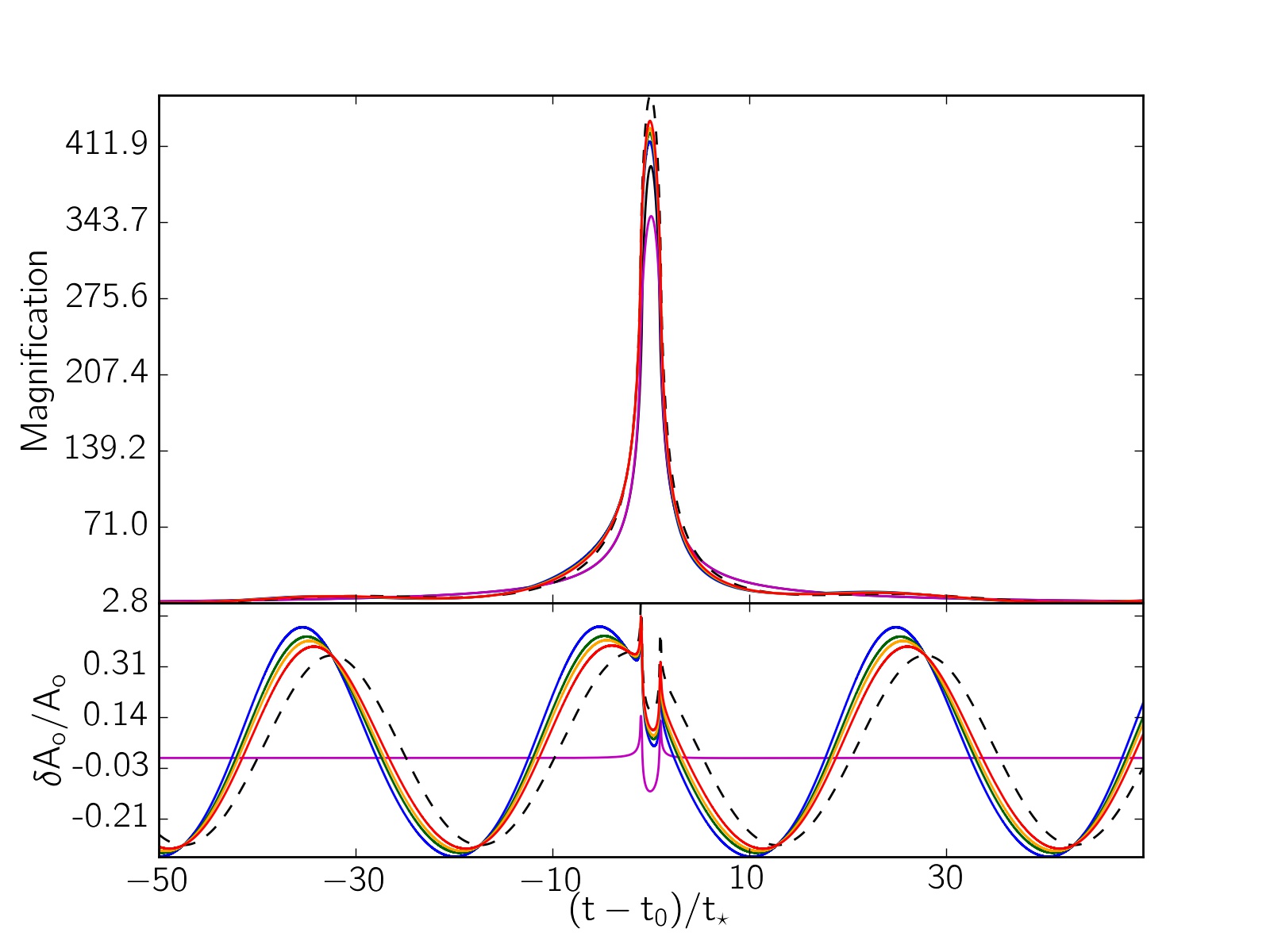}\label{figd}}
	\subfigure[]{\includegraphics[angle=0,width=0.49\textwidth,clip=0]{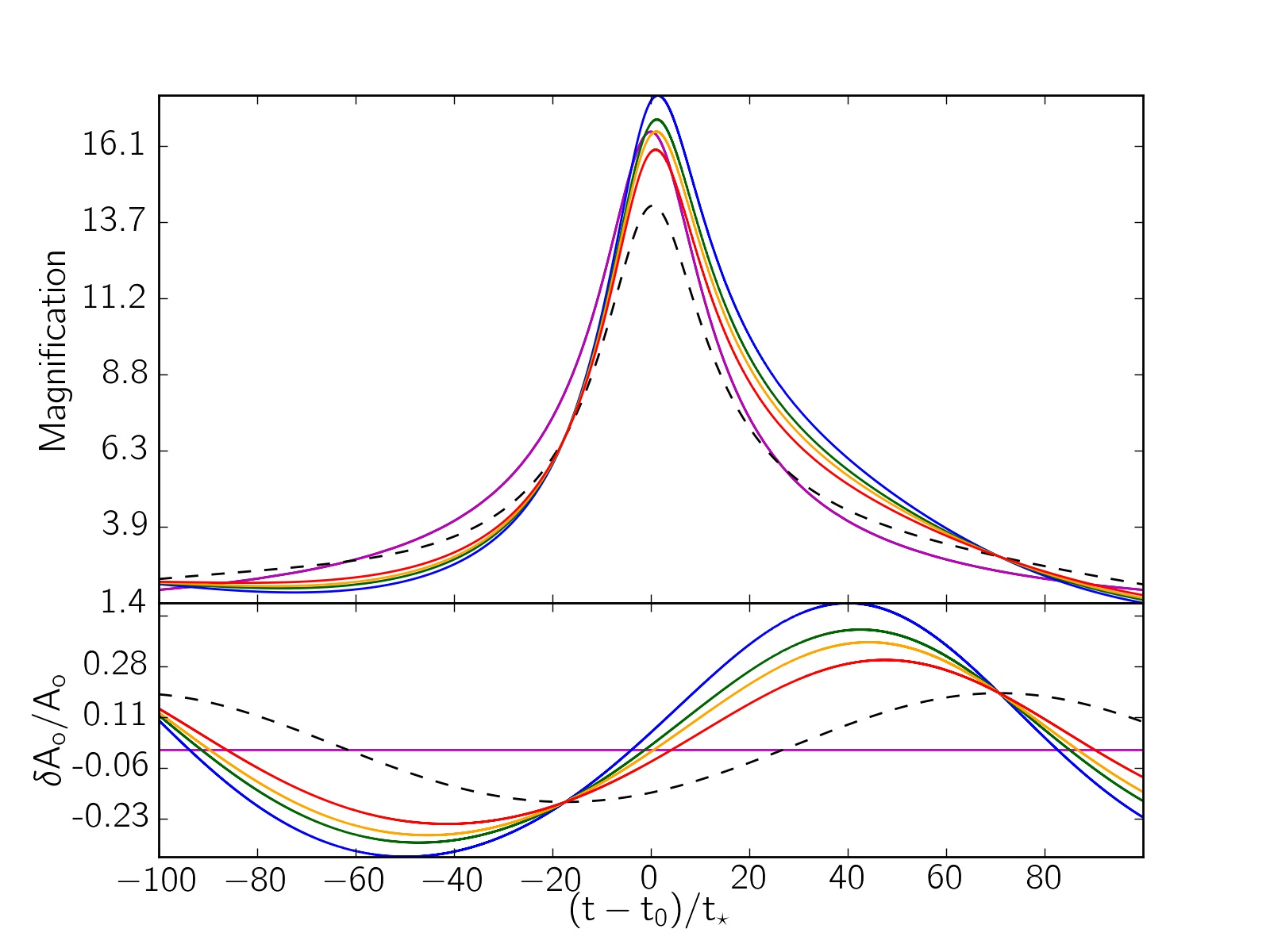}\label{fige}}
	\subfigure[]{\includegraphics[angle=0,width=0.49\textwidth,clip=0]{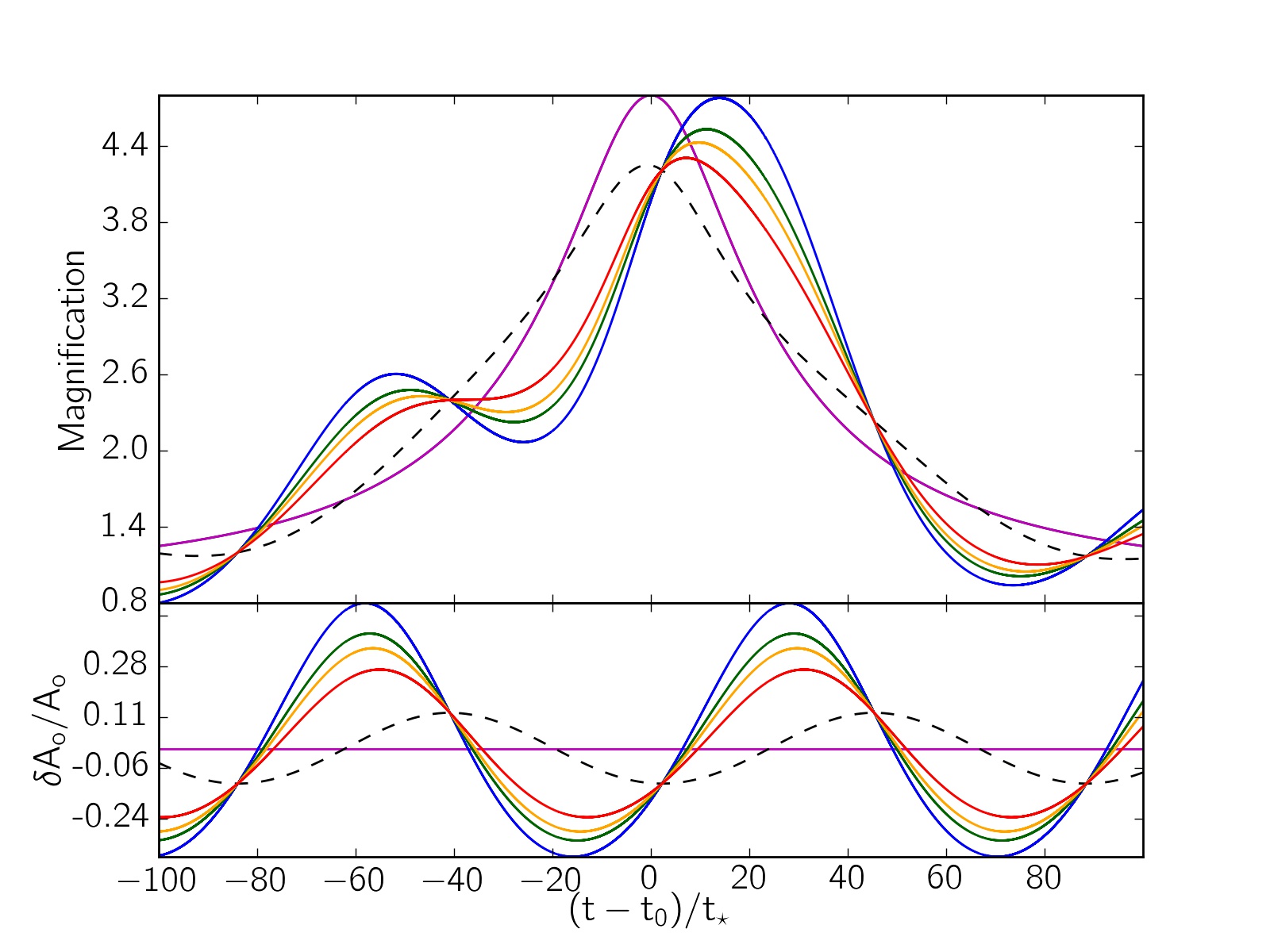}\label{figf}}
\caption{Examples of microlensing light curves of radial pulsating
stars. The parameters of each light curve can be found in each row of Table (\ref{table}). The
net microlensing light curve without pulsating effects are plotted
with solid black curve. The light curves with considering just
variation in $\rho_{\star}$ are shown in solid magenta curves.  The dashed black curves represent the microlensing light curves with applying the changes
in $\rho_{\star}$ as well as the factor of variation of source area, i.e., $A(\rho_{\star})(\frac{R_{\star}(t)}{\bar{R}})^{2}$. The curves of the observed magnification factor,
$A_{o}$, in $B$, $V$, $R$ and $I-$bands are represented by blue,
dark-green, orange and red solid curves, respectively. The residuals
$\delta A_{o}/A_{o}=(A_{o}(\rho_{\star})-A(\bar{\rho}_{\star}))/A(\bar{\rho}_{\star})$ are shown at the bottom of each light curve. }\label{light}
\end{figure*}
%%%%%%%%%%%%%%%%%%%%%%%%%%%%%%%%%%%%%%%%%%%%%%%%%%%%%%%
\begin{table*}
\centering
\caption{Table contains the parameters used to make microlensing light curves shown in Figures (\ref{light}) and (\ref{Binary}). }\label{table}
\begin{tabular}{ccccccccccc}
\hline
& $\delta_{R}$ & $\delta_{T}$  & $P$ & $t_{\rm{P}}$ &  $u_{0}$ & $\rho_{\star}$ & $t_{\rm{E}}$ & $\xi$& $d$ & $q$\\
  & $(R_{\star})$ & $\rm{(K)}$  & $\rm{(day)}$ & $\rm{(day)}$ &  $~$  & $~$ & $\rm{(day)}$ & $\rm{(deg)}$ & $(R_{\rm{E}})$ & $ $  \\
\hline
\ref{figa} & $0.30$  & $144.3$ &  $1.42$ &   $3.2$ &   $0.281$ &  $0.006$ &  $23.6$ &$-$ & $-$ & $-$ \\
\ref{figb} &  $0.33$ &  $240.9$ &  $1.90$ &   $-1.7$ &   $0.012$ &  $0.003$  & $18.3$  & $-$  &$-$&$-$\\%&  0.0527 \\ %% L3_51
\ref{figc}  & $0.30$ &  $223.1$ &  $7.48$ &   $2.9$ &     $0.029$ &  $0.003$ &  $23.3$  & $-$  &$-$& $-$  \\%0.0759
\ref{figd} &  $0.16$ &  $214.9$ &  $1.10$ &   $1.5$  &    $0.002$ &  $0.005$  & $7.6$   & $-$  &$-$& $-$\\%%% 0.0367  L4_0
\ref{fige} &  $0.09$ &  $382.7$ & $6.90$ &   $-0.7$ &    $0.060$  & $0.006$ &  $6.4$    & $-$ &$-$& $-$\\%  0.0391  L5_47
\ref{figf} &  $0.06$  & $381.7$ &  $5.90$  &  $0.3$ &     $0.210$  & $0.011$ &  $5.9$     & $-$ &$-$& $-$\\%  0.0684   L5_29
\ref{figBa} &$0.19$ & $172.7$ &  $2.31$ &  $23.9$  &   $0.062$ &  $0.013$ &  $56.5$   & $0.9$ & $0.89$ & $0.71$ \\
\ref{figBb}&  $0.32$ &  $107.9$ &  $12.88$ & $9.4$ &  $0.085$ &  $0.020$ &  $10.9$ & $17.5$ &  $0.68$ & $0.33$ \\
\ref{figBc} &  $0.24$ &  $383.2$ &  $4.63$ &  $6.5$ &   $0.006$ &  $0.053$  & $12.9$ & $-9.5$& $1.03$ & $0.73$ \\
\ref{figBd}& $0.13$ &  $285.7$ &  $4.11$ &  $23.9$ &   $0.354$  & $0.001$ &  $39.4$ & $268.2$ &  $0.74$ & $0.39$\\
\hline
\end{tabular}
\end{table*}
$(*)$: For the first plot, when the source radius is increasing (when
the $\phi$ changes from 0 up to 180), at times $0.25P$ and $0.75P$,
the lens distance is $u\sim \rho_{\star}$, corresponding to when the
maximum rates of deviations in the magnification factor occur. For these
events the maximum deviations in the magnification factor happens when
the lens is outside the source radius and the source size is minimum. For
a fixed $u$, decreasing and increasing $\rho_{\star}$ has an asymmetric
effect on the magnification factor.\\

$(*)$: Variations in $R_{\star}$ changes the projected source area with
significantly larger impact on the total magnification as compared to
variation in $\rho_{\star}$, even when $u \sim \rho_{\star}$.
Sign changes between the top and the middle panels are correlated.\\

$(*)$: Deviations in the observed magnification factor due to
$\delta_{T}\sim 200$~K and $\delta_{R}\sim 0.06\bar{R}$ are at
similar orders of magnitude (i.e., $0.1$). Likewise, pulsations with
$\delta_{T}=400$~K and $\delta_{R}=0.15\bar{R}$ produce similar levels of
variability in the observed magnification factor. However, our choice of
$\phi_0 = -\pi/2$ leads to a quarter shift in phase for the sign change
between the upper two panels and the lowest one for temperature.\\

$(*)$: According to the bottom plot, increasing the stellar temperature
produce a larger amplitude of deviation in the magnification as compared
to decreasing the stellar temperature. This occurs because in the Taylor
expansion for negative perturbation, the terms with odd power will be
negative whereas others are positive and will eliminate each other;
then for a positive perturbation, all powers have the same signs. \\

$(*)$: Increasing the stellar radius results in a larger deviation of the
magnification as compared to decreasing it
(see the middle plot). In considering the variational term for
$R_\ast$:

\begin{eqnarray}\label{del22}
\left(\frac{\delta A_{o}}{A_{o}}\right)_{2}=  \delta_{R}\, \sin (\omega (t-t_{p})+\phi_{0}) \,[2+ \sin(\omega(t-t_{p})+\phi_{0})].
\end{eqnarray}

\noindent The two terms in square brackets will add constructively if
$\sin[ \omega(t-t_{p})+\phi_{0}] >0$, but when the sinusoidal term is
negative, the amplitude of variation will be reduced.

In order to examine the overall perturbations in microlensing light curves due to stellar pulsation, we simulate the single and binary microlensing light curves from pulsing source stars  in the next sections. 

%%%%%%%%%%%%%%%%%%%%%%%%%%%%%%%%%%%%%%%%%%%%%%%%%%%%%%%%%%%%%%%%%
\section{Single-lens microlensing of pulsing stars}\label{three}

Figure~(\ref{light}) shows six simulated microlensing light curves of
radial pulsating stars. The parameters which have significant impacts
on the light curves are reported in Table (\ref{table}). Light
curves without pulsation are plotted with solid black curve (behind the
magenta curve). Light curves with only variations in $\rho_{\star}$ are
plotted as solid magenta; those that include variations of $\rho_{\star}$
and $(R_{\star}(t) / \bar{R})^{2}$ combined are shown as dashed
black curves.  Curves for the observed magnification factor $A_{o}$ in $B$,
$V$, $R$ and $I-$bands are shown by blue, dark-green, orange, and red
solid curves, respectively. The residuals with respect to the solid back
curves, $(A_{o}(\rho_{\star})-A(\bar{\rho}_{\star}))/A(\bar{\rho}_{\star})
= \delta A_{o}/A_{o}$, are plotted in the bottom panels of
each light curve.

We first assume that the lens impact parameter, i.e., $u_{0}$, is larger than the
source radius $(u_{0} > \bar{\rho}_{\star})$ (i.e., not transit of
the star by the lens).  In that case, microlensing only perturbs the
pulsation curve. The deviation in the variability due to microlensing
depends on the period and the width of the microlensing light curve.
If $ u_{0}t_{\textrm{E}} \gtrsim P$, several variable pulses magnify
due to lensing, as in Figure \ref{figa}. On the other hand, if $P$
is the same order as the light curve's width, one or two pulses will be
perturbed. In the case that the minimum of the intrinsic stellar intensity
happens at the time of closest approach, the deviation will be maximum
(see Figs~\ref{figb} and \ref{figc}).

Now we consider microlensing events with smaller lens impact parameter
$u_{0}\leq 10\rho_{\star}$. For these events, since the magnification
is high, the pulsation will produce perturbations on the main light
curve. Indeed, the general form of the microlensing light curve does
not change during lensing, so we interpret this as perturbation on the
microlensing light curve due to pulsations. According to the previous
section, the variability of $\rho_{\star}$ will be significant only for
transit microlensing events ($\rho_{\star} \gtrsim u_{0}$). In that case,
the solid magenta curves which isolate the effect of the magnification
factor dependence on $\rho_{\star}$ as a function of time will be deviate
from solid black curve, as shown in Figure \ref{figd}, otherwise the
solid magenta and dashed black curves are always coincident.  This point
can be inferred from the right panel of Figure~(\ref{delta1}). In
Figure~\ref{figd} because of high magnification factor, the light
curves in different filters are similar.

We note that in the analysis of the first microlensing
event of a pulsating source star reported by \citet{Varmicro},
the effect of a varying $\rho_{\star}$ was not considered. In this event,
although the pulsation period which was estimated at $\sim 9~$days is longer
than $t_{\star}\sim3.7~$days, the lens was crossing the source surface
($u_{0}<\rho_{\star}$). Consequently, variation
in $\rho_{\star}$ alters the slope
of the light curve and its peak value. By fixing $\rho_{\star}$,
a wrong estimate for the lens impact parameter could result.

If the source is at neither maximum nor minimum at the time of closest
approach between the lens and the source center, the two sides of the
microlensing light curve will not be symmetric with respect to $t_{0}$.
Figures \ref{fige} and \ref{figf} emphasize the resulting asymmetry that
can result.

When $\delta_{T}$ is large ($\gtrsim 200$~K), the difference between light
curves for different filters can be significant (e.g., compare Figs.\
\ref{figc} and \ref{fige}).  Generally, in the microlensing of pulsating
stars, the time of the maximum magnification will not be exactly $t_{0}$,
unless the peak of the source intrinsic brightness happens exactly at the
time of the closest approach (which is rare). And the modified maximum
times are not the same in different filters. Since the perturbation on the
light curve due to pulsation is higher for shorter wavelengths, the light
curves in the $B-$band will show a greater displacement in the time of
the magnification peak (from $t_{0}$) as compared with other filters. This
point is shown in Figures~\ref{fige} and \ref{figf}. The phase difference
between the residual curves (dashed black ones and colored solid ones)
vanishes as $\delta_{T}$ decreased.  For large values of $\delta_{T}$
(e.g.,  $\gtrsim 200$~K),  the phase differences are enhanced.
\begin{figure*}
\centering
\subfigure[]{\includegraphics[angle=0,width=0.49\textwidth,clip=0]{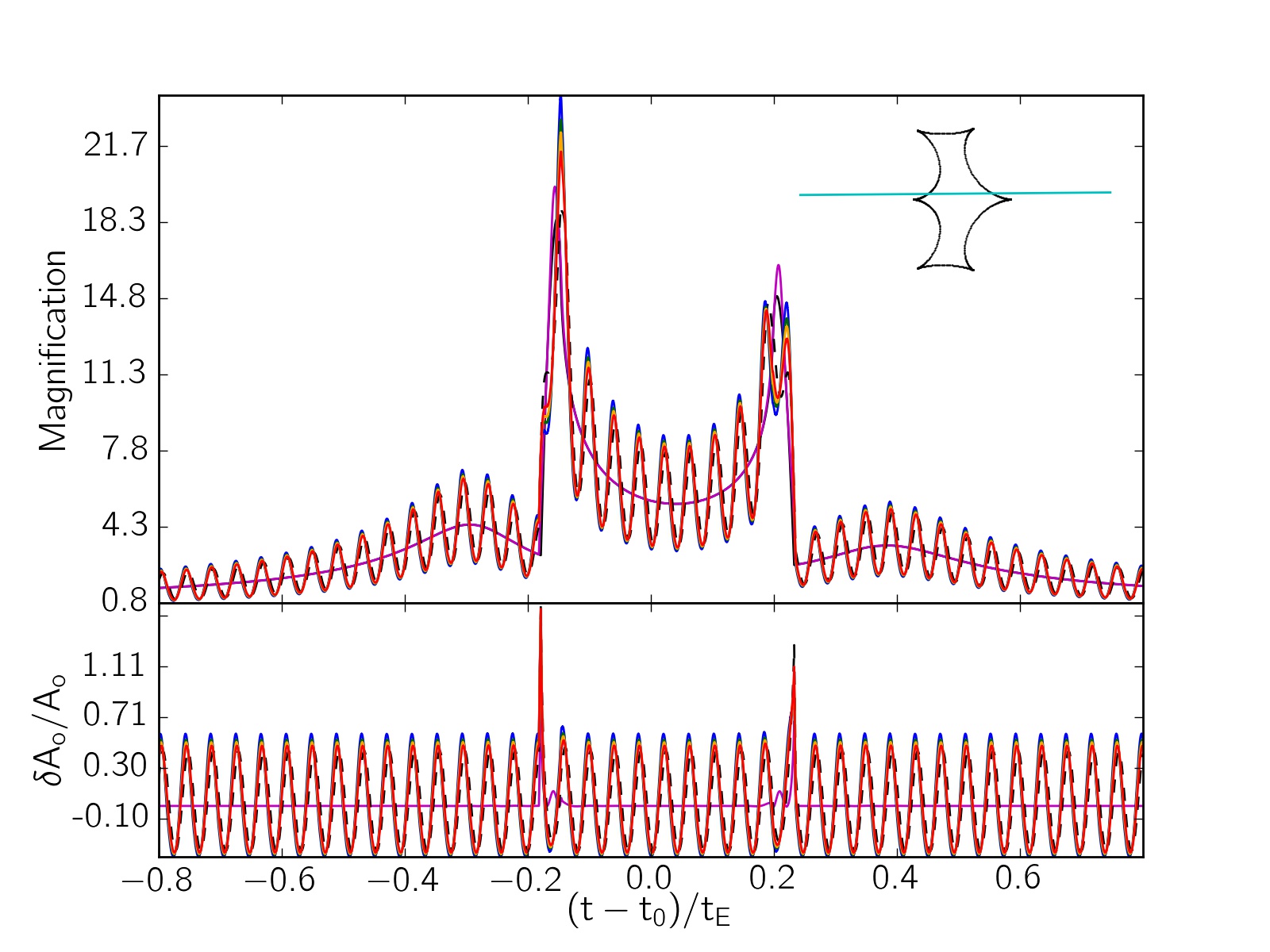}\label{figBa}}
\subfigure[]{\includegraphics[angle=0,width=0.49\textwidth,clip=0]{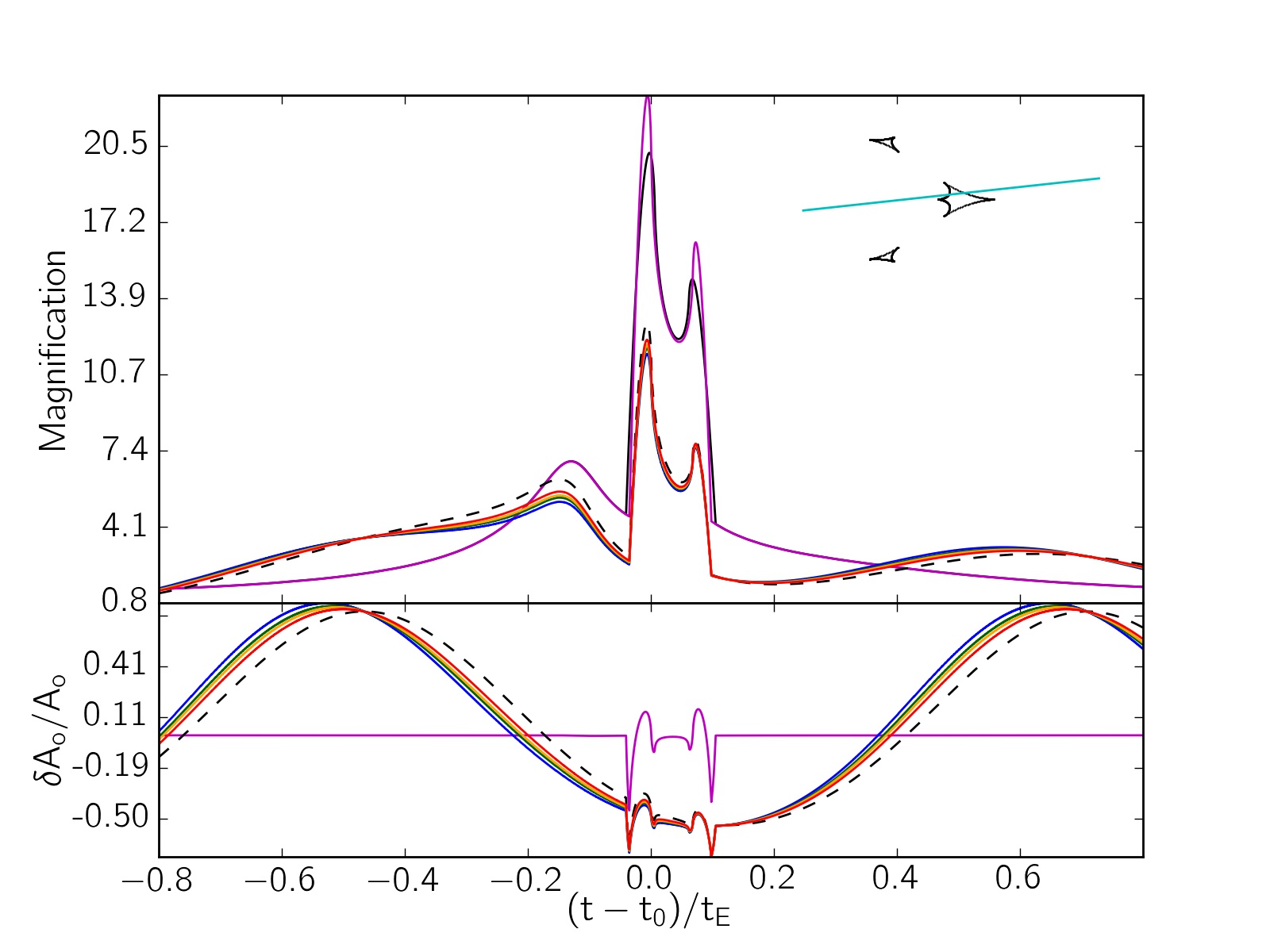}\label{figBb}}
\subfigure[]{\includegraphics[angle=0,width=0.49\textwidth,clip=0]{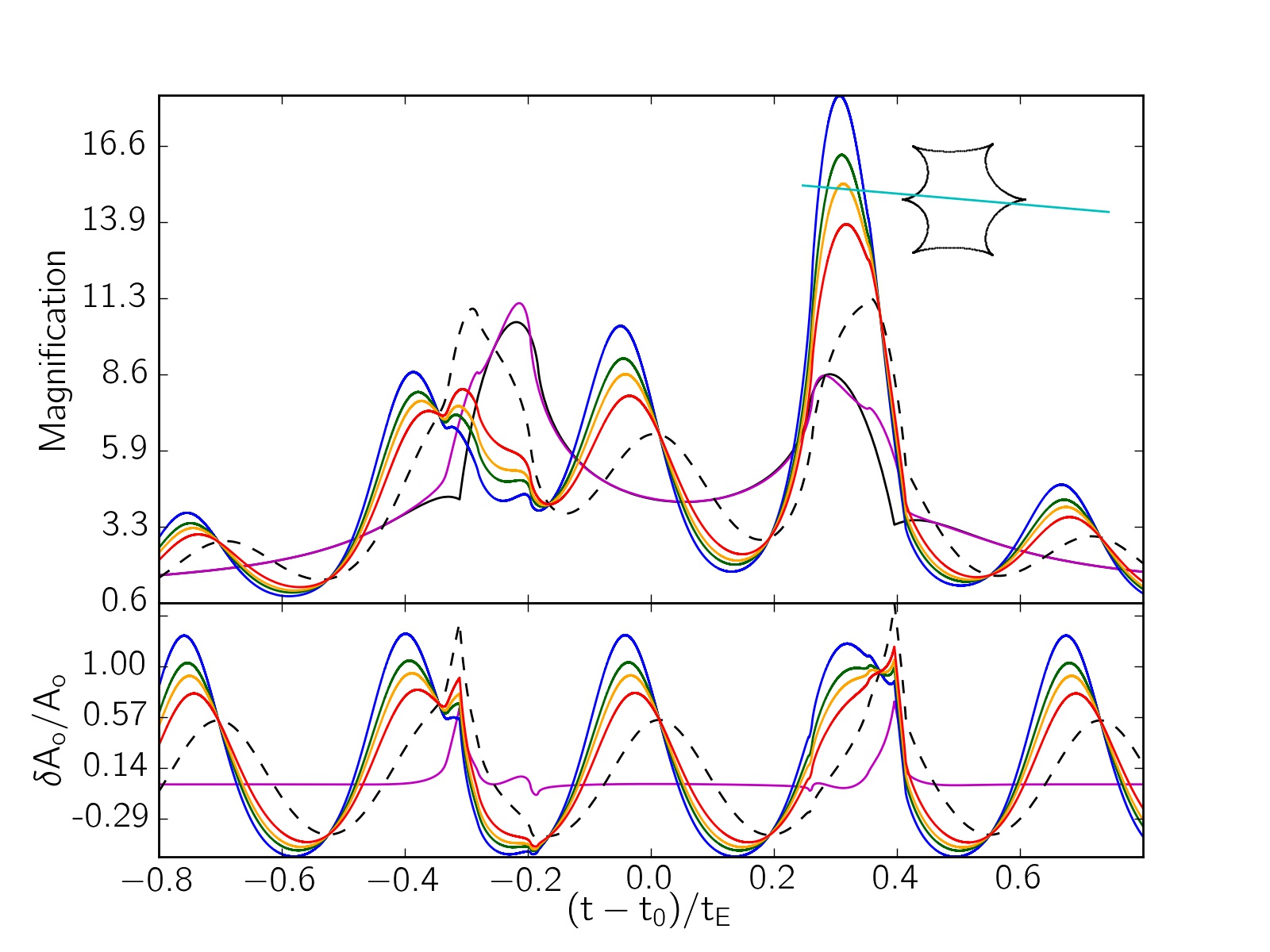}\label{figBc}}
\subfigure[]{\includegraphics[angle=0,width=0.49\textwidth,clip=0]{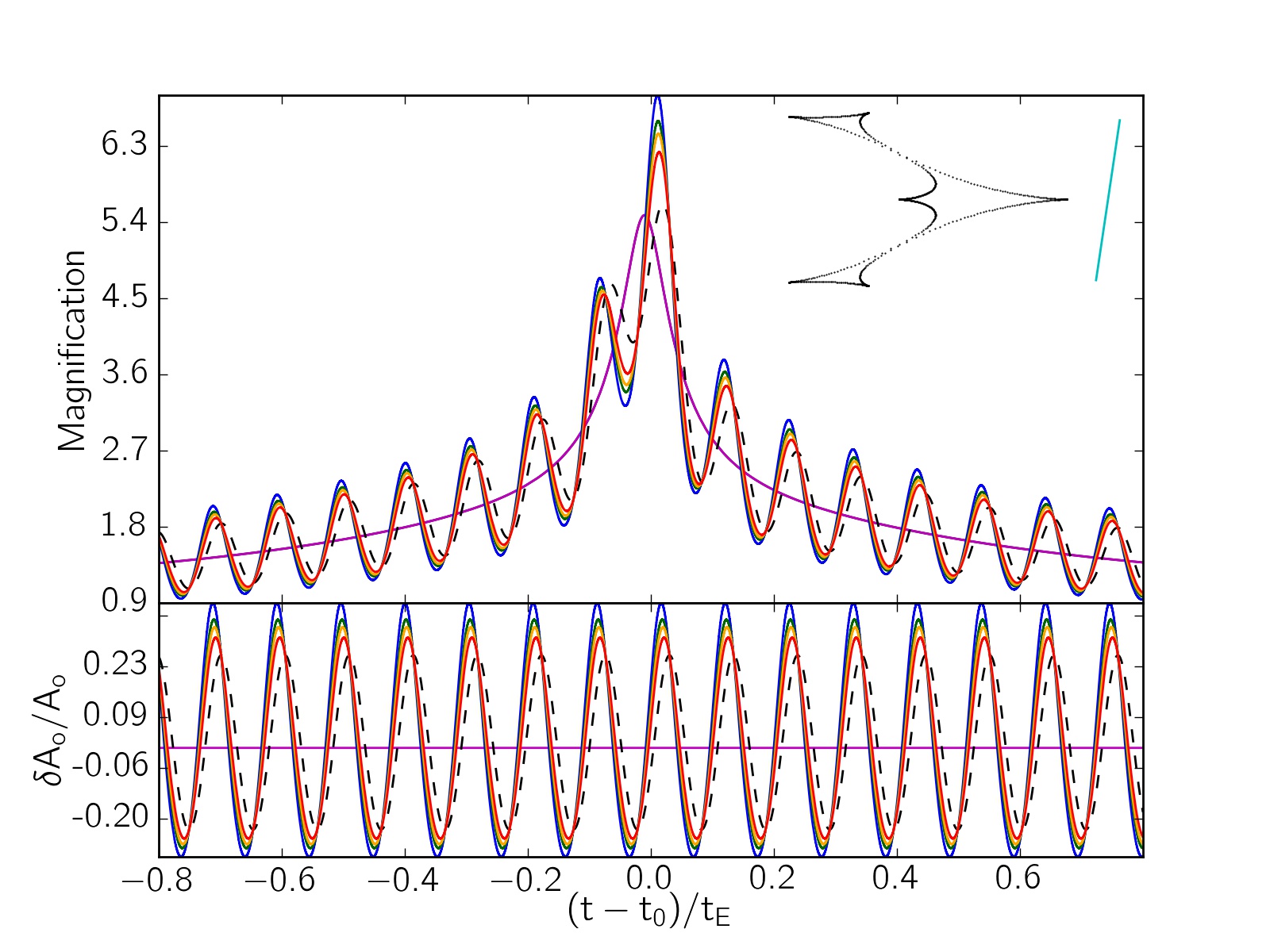}\label{figBd}}
\caption{Examples of binary microlensing light curves from pulsing source stars. The parameters used to make each of them are listed in Table(\ref{table}). More information about the plots can be found in the caption of Figure (\ref{light}).}\label{Binary}
\end{figure*}
   
%%%%%%%%%%%%%%%%%%%%%%%%%%%%%%%%%%%%%%%%%%%%%%
\section{Binary microlensing of pulsing stars}\label{four}

Among reported microlensing events, a remarkable fraction involve binary
lenses with caustic-crossing features \citep{Wyrzykowski2015b,alcock2000}.
When a background source has a caustic crossing, the magnification
factor enhances significantly so that small perturbations on the
stellar surface or in its flux can be greatly magnified and detectable
\citep[see, e.g.,][]{schneider1992,Gaudi2012}. Here, we study if the
pulsing signatures from source stars can be revealed in caustic crossing
features, or when the source stars pass close to the caustic curves.

Figure (\ref{Binary}) shows four examples of simulated binary microlensing
events involving pulsing source stars. Their characterizations are
the same as Figure~(\ref{light}), with model parameters provided in
Table~(\ref{table}). The three last columns of the Table are relevant
for binary lenses but not single lenses: $\xi$ is the angle of source trajectory relative to the binary axis in the lens plane, 
$d(R_{\rm{E}})$ is the binary
distance normalized to the Einstein radius, and $q$ is the ratio of the
two stellar masses, respectively.  Here we summarize several key
aspects of the simulations.

\begin{figure}
\centering
\includegraphics[width=0.49\textwidth]{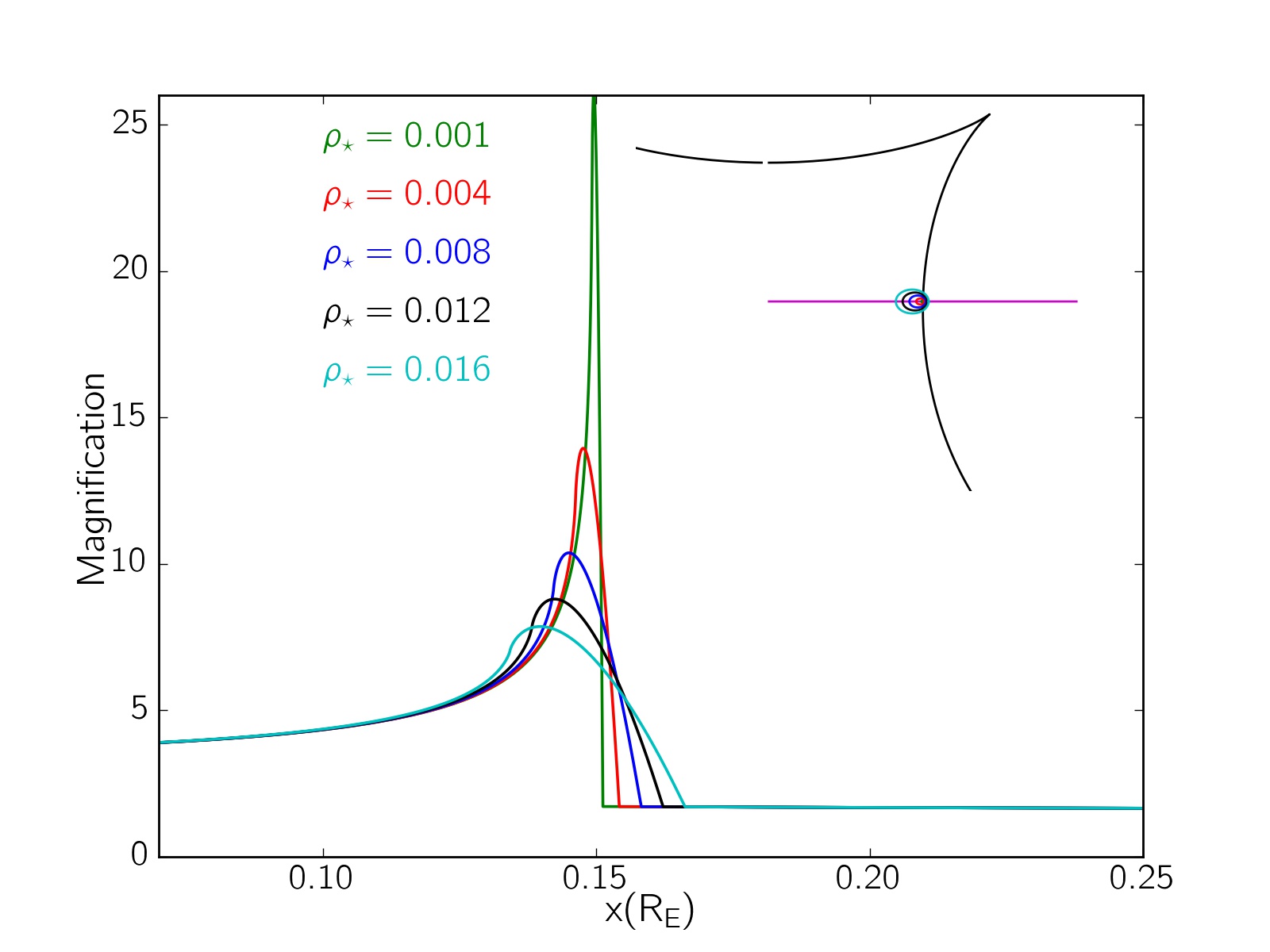}
\caption{The magnification factor in binary caustic crossing versus the horizontal distance of the source center from the lenses' center of mass for several values of the source radius. The parameters used to make this plot are:  $q= 1.0$, $d(R_{\rm{E}})=1.0$ and $u_{0}=0.4$.  The source trajectory is parallel with the binary axis. }\label{fold}
\end{figure} 

(i) In binary microlensing from variable stars, the strongest effects
occur during caustic crossing, owing to the variation in the finite source
size, as emphasized with Figures \ref{figBa}  and \ref{figBb}. In residual
panel, some deviations in the magenta solid curves develop right at the
times of caustic crossings. The durations of these deviations are
short, on the order of $t_{\star}$ for how the source takes to transit
the caustic line. If $P \gg t_{\star}$, the star is essentially of fixed
angular extent during the caustic crossing.  In that case discerning
the periodic variability of the source radius during the crossing is
difficult. If $P\sim t_{\star}$ (either short period pulsations, or low lens-source relative proper motion, or large finite source size), then
the effect of source variation is significant and can be discerned.
Note that observational sampling of the light curve should be much
shorter than $t_{\star}$.  Certainly, the detection of variability in
the source radius is more likely when $t_{\star}$ is long.

(ii) While fold crossing, the position of the maximum magnification
depends on the source size. This point is demonstrated in
Figure~(\ref{fold}).  The brightness peak during a fold-crossing changes
position in response to the pulsational variation of the star, even for
the magenta solid curves. When the source radius decreases, the peak
becomes more slender and moves toward the fold line, and vise versa.

(iii) Generally the pulses of source stars have larger amplitudes while
the magnification factor is high, as for example when the source is
inside the caustic curve or when the source is quite close to the cusp
of the caustic (see Figure \ref{figBa}).

(iv) When the source is passing either close to the caustic line
or tangential to it, an extended peak develops in light curve. Under
these conditions, the variation of finite source size does not produce
particularly remarkable deviations from a non-pulsing star (e.g., the
mangeta solid curves are over the solid black curves). One example is
shown in Figure \ref{figBb} around the time $0.2t_{\rm{E}}+t_{0}$.

(v) If the pulsing amplitude is large $\delta_{T}\sim 300~$K,  the
enhancements in the apparent brightness due to variability and lensing
can be in the same order of magnitudes. In that case, the resulted light
curve will be complicated, see, e.g.  Figure \ref{figBc}.  In these
light curves, a significant displacement in the maximum magnification
happens from the time of caustic-crossing.

(vi) In the absence of a caustic crossing (e.g., Fig.~\ref{figBd}),
the deviation due to variable finite source size is negligible and other
deviations are similar to single microlensing events with large impact
parameter, such as magnifying several of the pulses.

%%%%%%%%%%%%%%%%%%%%%%%%%%%%%%%%%%%%%%%%%%%%%%%%
\section{Summary and conclusions}

Gravitational lensing has proven to be a powerful tool for study of
the universe, ranging from macrolensing effects that have been used
to study dark matter in galaxies and galaxy clusters, to microlensing
phenomena for constraining dark matter candidates in the Milky Way halo.
But microlensing has also been used as a probe of stellar astrophysics.
For a close passage between a lensing mass and a background stellar
source, differential magnification across the stellar source allows for a
version of mapping the star.  While this potential has been investigated
theoretically for the effects of limb darkening, atmospheric polarization,
and even circumstellar media, much less has been done to address the
effects of stellar pulsation.

In this paper we consider specifically radially pulsating stars in the
context of single-lens and binary-lens events.  We adopt a simplified
model for the pulsational behavior, as a sinusoidal variation in the
radius and in the temperature (with a phase difference).  For this
initial study, we ignored limb darkening.  For the
single-lens case, we exploit this simple picture to deconstruct how
different factors contribute to lensing light curves, including
how the surface-averaged magnification factor varies with the
changing source size, how the intrinsic source brightness
changes with sources size, and how the stellar intensity changes
with temperature variation.  

The main outcomes from simulations of the single-lens models are
as follows. Certainly when finite source effects are not important,
the brightness variations from pulsational variability are simply
mimicked in the lensing light curve.  However, when the lens
is quite near or transits the source, interesting effects occur,
such as time-lag and amplitude differences for different passbands.
Details regarding the strength of these effect depend on how
the pulsation period, $P$, compares with duration of the lensing
event, $t_E$, as well as how the pulsational phase (i.e., maximum or
minimum brightness) compares to when the lens achieves its closest
approach.

Binary lensing yields similar effects, although the magnification factor
is more complex.  The light curves in general now depend on the mass
ratio of the lens, the binary separation, and the relative trajectory of
the source in relation to the binary lens caustic pattern.  As with the
single-lens case, effects depend on the how the pulsation period compares
with the duration of the lensing.  Likewise, the observed brightness
variation from pulsation can be strongly modified as compared to the
unlensed case, and different curves can result for different passbands.
For a single lens, the possibility of finite source effects is higher for
stars of larger size, and small impact parameters are needed.  For binary
lenses the caustic structure is relatively larger, and caustic crossings
always lead to high magnification effects, so finite source events are
easier to detect for binary lenses.

An interesting distinction about the case of radial pulsators, as compared
with finite source effects with non-pulsating stars, is the possibility
of obtaining a good measurement for the source distance.  When dealing
with Galactic microlensing events, there can be degeneracies, such as
the mass of the lens, the distance of the lens, the trajectory of the
relative proper motion between the lens and source, and to some extent
the distance to the source.  The latter can be somewhat constrained.
For example, in monitoring surveys of the LMC, the relative error for
the source distance can be small, given that the LMC is much farther
than its size.  By contrast, there can be more ambiguity for the
source distance when monitoring the Galactic Bulge.  However, radial
pulsators such as Cepheids and RR Lyrae stars are standard candles that
follow a Leavitt Law relation for period of pulsation and luminosity.
Consequently, microlensing that involves such a stellar source will
provide an accurate determination of $D_{\rm s}$.

In a second paper, we will explore the effects of non-radial
pulsational (NRP) brightness variations on microlensing light curves.
Overall, we can expect the amplitudes of intrinsic brightness 
variations to be lower than for radial pulsators.  On the other hand, 
finite source effects from transits of the source by the lens
should permit a determination of the trajectory of the transit
relative to the principle axis of the star that defined the 
spherical harmonic $l,m$ modes associated with the NRP variations.
We also anticipate some sensitivity to stellar rotation in
favorable cases.

\section*{Acknowledgements}

RI acknowledges Ashton Morelock for preliminary calculations of single
lens light curves of variable stars.

\bibliographystyle{mnras}
\bibliography{paper}
\end{document}